\documentclass[journal]{IEEEtran}
\IEEEoverridecommandlockouts    
\usepackage{graphicx}
\usepackage{subfigure}
\usepackage[utf8]{inputenc}
\usepackage{amsfonts,amssymb,amsmath}
\usepackage{stix}
\usepackage{caption}
\usepackage{color}
\usepackage[hyphens]{url}
\usepackage[hidelinks]{hyperref}
\usepackage{array}
\usepackage{multirow}
\usepackage{bigstrut}
\usepackage{tabularx}
\usepackage{makecell}
\newtheorem{insight}{\bf Insight}
  
\newcommand{\ignore}[1]{}
\usepackage[T1]{fontenc}
\newcommand{\V}{{{\cal V}}}
\newcommand{\A}{{{\cal A}}}
\newcommand{\D}{{{\cal D}}}
\newcommand{\myfont}{\fontfamily{lmtt}\selectfont\normalsize}

\usepackage{listings, xcolor}
\definecolor{verylightgray}{rgb}{.97,.97,.97}

\lstdefinelanguage{Solidity}{
	keywords=[1]{anonymous, assembly, assert, break, balance,  callcode, case, catch, class, constant, continue, constructor, contract, debugger, default, delegatecall, delete, do, else, emit, event, experimental, export, external, finally, for, function, gas, if, implements, import, in, indexed, instanceof, interface, internal, is, length, library, log0, log1, log2, log3, log4, memory,mapping, modifier, new, pragma, private, protected, public, pure, push, require, returns, revert, selfdestruct, send, solidity, storage, struct, suicide, super, switch, then, this, throw, try, typeof, using, view, while, with, addmod, ecrecover, keccak256, mulmod, ripemd160, sha256, sha3,Transfer}, % generic keywords including crypto operations
	keywordstyle=[1]\color{blue},
	keywords=[2]{address, bool, byte, bytes, bytes1, bytes2, bytes3, bytes4, bytes5, bytes6, bytes7, bytes8, bytes9, bytes10, bytes11, bytes12, bytes13, bytes14, bytes15, bytes16, bytes17, bytes18, bytes19, bytes20, bytes21, bytes22, bytes23, bytes24, bytes25, bytes26, bytes27, bytes28, bytes29, bytes30, bytes31, bytes32, enum, int, int8, int16, int24, int32, int40, int48, int56, int64, int72, int80, int88, int96, int104, int112, int120, int128, int136, int144, int152, int160, int168, int176, int184, int192, int200, int208, int216, int224, int232, int240, int248, int256, string, uint, uint8, uint16, uint24, uint32, uint40, uint48, uint56, uint64, uint72, uint80, uint88, uint96, uint104, uint112, uint120, uint128, uint136, uint144, uint152, uint160, uint168, uint176, uint184, uint192, uint200, uint208, uint216, uint224, uint232, uint240, uint248, uint256, var, void, ether, finney, szabo, wei, days, hours, minutes, seconds, weeks, years},	% types; money and time units
	keywordstyle=[2]\color{teal},
	keywords=[3]{block, blockhash, coinbase, difficulty, gaslimit, number, timestamp, data, gas, sig, now, tx, gasprice, origin, call},	% environment variables
	keywordstyle=[3]\color{violet},
	identifierstyle=\color{black},
	sensitive=false,
	comment=[l]{//},
	morecomment=[s]{/*}{*/},
	commentstyle=\color{black},
	stringstyle=\color{red},
	morestring=[b]',
	morestring=[b]"
}

\begin{document}
\title{A Survey on Ethereum Systems Security: \\ Vulnerabilities, Attacks and Defenses}
\author{Huashan Chen, Marcus Pendleton, Laurent Njilla, and Shouhuai Xu
\thanks{H. Chen and S. Xu are with the Department of Computer Science, University of Texas at San Antonio. 
M. Pendleton and L. Njilla are with the U.S. Air Force Research Laboratory, Rome, NY.
Correspondence: {\tt shxu@cs.utsa.edu}}}
\maketitle

\begin{abstract}
The blockchain technology is believed by many to be a game changer in many application domains, especially financial applications. While the first generation of blockchain technology (i.e., Blockchain 1.0) is almost exclusively used for cryptocurrency purposes, the second generation (i.e., Blockchain 2.0), as represented by Ethereum, is an open and decentralized platform enabling a new paradigm of computing ---  Decentralized Applications (DApps) running on top of blockchains. The rich applications and semantics of DApps inevitably introduce many security vulnerabilities, which have no counterparts in pure cryptocurrency systems like Bitcoin. Since Ethereum is a new, yet complex, system, it is imperative to have a systematic and comprehensive understanding on its security from a holistic perspective, which is unavailable. To the best of our knowledge, the present survey, which can also be used as a tutorial, fills this void. In particular, we systematize three aspects of Ethereum systems security:  vulnerabilities, attacks, and defenses. We draw insights into, among other things, vulnerability root causes, attack consequences, and defense capabilities, which shed light on future research directions.
\end{abstract}


\IEEEpeerreviewmaketitle

\section{Introduction}
The notion of {\em blockchain} was implicitly introduced in 2008 as the key underlying technique of the cryptocurrency known as Bitcoin \cite{nakamoto2008bitcoin}, which uses a {\em transaction-centered} model known as {\em unspent transaction outputs} (UTXO). In this model, a blockchain is a distributed and public ledger, which records the payment transactions between parties over a peer-to-peer (P2P) network. Unlike traditional digital cash systems \cite{DBLP:conf/crypto/Chaum82}, in which there is a trusted third party (e.g., bank), there is no trusted third party in a blockchain system in general and in Bitcoin in particular. Bitcoin is often referred to as Blockchain 1.0 because it only offers payment services. The innovation of the Bitcoin system is its consensus protocol, which allows mutually distrusting nodes in a P2P network to eventually reach a consensus on the outcome after executing payment transactions. Unlike traditional consensus protocols \cite{DBLP:journals/jacm/FischerLP85}, the participants are from an open network and are incentivized by the payment of Bitcoins (or BTCs), which are ``mined'' through a clever cryptographic hash function known as Proof-of-Work (PoW), an idea originally proposed as an anti-spam technique \cite{DBLP:conf/crypto/DworkN92}.

Perhaps inspired by the success of Bitcoin as well as the need to support semantically richer (than just payment) applications, the notion of {\em smart contracts} has been introduced to represent autonomous programs, leading to a new paradigm of Decentralized Applications (DApps) that run on top of blockchains and consist of many interacting smart contracts. The Ethereum system was launched in 2015 to support smart contracts, while offering its inherent cryptocurrency known as Ether \cite{wood2014ethereum} and using an {\em account-centered} model (rather than the UTXO model mentioned above). Ethereum has become the de facto standard platform for DApps. At the moment of writing, the market value of Ethereum is over US\$31B with approximately one million smart contracts executing on top of the Ethereum blockchain \cite{Ethereum49:online, SmartCon86:online}.
The success of Ethereum ushers in Blockchain 2.0, which goes much beyond the payment-centered Blockchain 1.0.

While Ethereum facilitates semantically richer applications than Bitcoin, it also enlarges the threat surface, as evidenced by the many high-profile attacks. One example is the DAO \cite{TheDAO} attack in year 2016, in which case an attacker exploited the so-called reentrancy vulnerability (which will be detailed later) to steal approximately US\$60M. In July 2017, a vulnerability in the Parity wallet contract caused the loss of US\$31M \cite{Parity}. 
In April 2018, the MyEtherWallet wallet fell victim to a BGP and DNS hijacking attack, enabling the hacker to steal approximately US\$17M \cite{MyEtherWallet}. These attacks highlight that our capabilities in securing the Ethereum system are limited. This should not be taken as a surprise because Ethereum is a new programming paradigm with DApps running on top of blockchains with many autonomous contracts. 

The motivation of the present survey is in three-fold of {\em researchers}, {\em practitioners}, and {\em students}. From the standpoint of a {\em researcher} who wants to investigate Ethereum security, there is a need for a source of systematized treatment on the problems related to Ethereum security. Despite the fact that there have been some surveys, they did not offer a systematic and comprehensive view on Ethereum vulnerabilities, attacks, and defenses as we do. While referring to the related prior work in Section \ref{sec:related-work} for details, we mention the following: there is neither systematic understanding of the Ethereum vulnerabilities that have been discovered, nor systematic understanding of their root causes; this may explain why there are still a number of vulnerabilities that are completely open. 
From the standpoint of a {\em practitioner}, there is a need for a source of best practices and guiding principles. Industry has conducted due diligence in summarizing many best practices \cite{Solidity80:online}, which however may overwhelm practitioners. Therefore, it might be more useful to have a small number of guiding principles that are easier to adopt in practice.
From the standpoint of a {\em student} who wants to learn about Ethereum security, there is a need for a succinct yet comprehensive and systematic source that also offer references to materials of greater details.

\subsection{Our contributions} 
We provide a systematic and comprehensive survey on Ethereum systems security. It is systematic in the sense that {\em vulnerabilities}, {\em attacks}, and {\em defenses} as well as the relationships between them are accommodated. It is comprehensive in the sense that it covers both the Ethereum platform via a layered architecture and the environment in which the Ethereum platform operates. 
In terms of vulnerabilities, we enumerate 44 types of Ethereum vulnerabilities according to the layers of the Ethereum architecture and the environment in which Ethereum operates. Perhaps more importantly, we analyze and systematize the root causes of those vulnerabilities. This allows us to provide insights into how to prevent some Ethereum vulnerabilities and how to cope with the inevitable vulnerabilities, including those that are largely open.
Some of our findings and insights are highlighted as follows:
\begin{enumerate}
\item Ethereum smart contracts introduce new kinds of vulnerabilities that do not have counterparts in traditional paradigm of applications.
\item It is important to design more secure programming languages and supporting tools for programmers to write more secure smart contracts.
\item The vulnerabilities caused by the design and implementation of the Ethereum blockchain are harder to cope with than vulnerabilities in the traditional paradigm.
\item The vulnerabilities in the Ethereum environment are largely caused by human, usability, and networking factors.
\item There are many vulnerabilities (e.g., {\em outsourceable puzzle}, {\em 51\% hashrate}, and {\em under-priced opcodes}) that must be tackled in order to adequately defend Ethereum or blockchain-based DApp systems in general.
\end{enumerate}

In terms of attacks, we systematize 26 attacks against Ethereum according to the layers of the Ethereum architecture. Perhaps more importantly, we relate these attacks to the vulnerabilities and systematize the attack consequences. Some of our findings are highlighted as follows:
\begin{enumerate}
\item The largest single-incident financial loss thus far was caused by a denial-of-service (DoS) attack against the Parity wallet. This attack disabled a library that is used by many contracts and is therefore quite different from traditional DoS attacks.
\item DApps running on top of blockchains are not fully decentralized when they use centralized web interfaces, even if the underlying blockchain is fully decentralized.
\item Ethereum application-layer attacks have caused the largest financial losses.
\end{enumerate}

In terms of defenses, we systematize 47 defenses into two classes: proactive defenses, which aim to prevent attacks as much as possible; and reactive defenses, which aim to cope with the hidden vulnerabilities (i.e., their existence may not be known to the defender). 
Perhaps more importantly, we present a deeper analysis of the defenses according to (i) their capabilities in mitigating the damages that can be incurred by exploiting vulnerabilities, and (ii) the defense effort that has been investigated to cope with each kind of vulnerabilities. Some of our findings are highlighted as follows:
\begin{enumerate}
\item Industry has come up with a significant set of best practices for guiding the development of smart contracts. These best practices, if adequately executed, can indeed avoid many vulnerabilities.
\item Existing proactive defenses can defend against attacks that exploit many vulnerabilities; in contrast, existing reactive defenses can only defend against attacks that exploit a few vulnerabilities.
\item There are large discrepancies between levels of effort and investment into different kinds of attacks (i.e., much more investment into defending against high-profile attacks than low-profile attacks).
\item Existing studies focus on defending against attacks that attempt to exploit vulnerabilities in the DApp back-end (i.e., smart contracts), but largely ignore the protection of the DApp front-end (i.e., browser) and the interactions between the front-end and the back-end.
\end{enumerate}

Although the present paper focuses on the Ethereum system, the aforementioned findings related to vulnerabilities, attacks and defenses might be applicable to blockchain-based systems in general. Moreover, we discuss some fundamental research problems that must be adequately tackled in order to secure and defend Ethereum and blockchain-based systems, including:
\begin{enumerate}
\item There is a lack of deep understanding on the rigorously-defined properties of blockchain-based systems.
\item There is a lack of deep understanding on the rigorous analysis methodologies that are necessary and sufficient for analyzing the desired properties of blockchain-based systems.
\item There is a lack of deep understanding on the metrics that are necessary and sufficient for analyzing the security and risk of  blockchain-based systems.
\end{enumerate}

In order to improve readability and ease the reference to the large number of vulnerabilities, we denote the vulnerabilities by $\V_1,\ldots,\V_{44}$, respectively. Similarly, we respectively denote the attacks by $\A_1,\ldots,\A_{26}$ and the defenses by $\D_1,\ldots,D_{47}$.
Moreover, we use ``${\cal A}_i({\cal V}_j,\cdots)$'' to denote that attack ${\cal A}_i$ exploits vulnerabilities ${\cal V}_j$ and possibly others.

The preceding discussion justifies how the present paper would adequately serve the needs of students and researchers.
For practitioners, we mention that we systematize the best practices into a small number of principles that may be easier to adopt in practice.

\subsection{Related work} 
\label{sec:related-work}

\begin{table*}[!htbp]
  \centering
      \scriptsize
  \setlength\tabcolsep{2pt}
  \caption{Comparison between surveys related to blockchain security and privacy.}
    \begin{tabular}{|c|c|c|c|c|c|c|c|c|c|c|c|}
    \hline
    \multirow{2}[4]{*}{Study} & \multicolumn{3}{c|}{Vulnerabilities} & \multicolumn{3}{c|}{Attacks} & \multicolumn{3}{c|}{Defenses} & \multicolumn{2}{c|}{Building-blocks} \bigstrut\\
\cline{2-12}      & \multicolumn{1}{p{6em}|}{Blockchain 1.0} & \multicolumn{1}{p{6em}|}{Blockchain 2.0} & \multicolumn{1}{p{3em}|}{Insights} & \multicolumn{1}{p{6em}|}{Blockchain 1.0} & \multicolumn{1}{p{6em}|}{Blockchain 2.0} & \multicolumn{1}{p{3em}|}{Insights} & \multicolumn{1}{p{6em}|}{Blockchain 1.0} & \multicolumn{1}{p{6em}|}{Blockchain 2.0} & \multicolumn{1}{p{3em}|}{Insights} &
\multicolumn{1}{p{5.5em}|}{Cryptography} &
\multicolumn{1}{p{4.1em}|}{Consensus} \bigstrut\\
    \hline
    \cite{atzei2017survey} &   & \multicolumn{1}{c|}{12} &   &   & \multicolumn{1}{c|}{9} &   &   & 3 &   &   &  \bigstrut\\
    \hline
    This work &   & \multicolumn{1}{c|}{44} & \footnotesize{\checkmark}  &   & \multicolumn{1}{c|}{26} & \multicolumn{1}{c|}{\footnotesize{\checkmark}
} &   & 47  & \footnotesize{\checkmark}
 &   &  \bigstrut\\
    \hline \hline
    \cite{LI2017} & \multicolumn{2}{c|}{20} &   & \multicolumn{2}{c|}{6} &   & \multicolumn{2}{c|}{5} &   &   &  \bigstrut\\
    \hline
    \cite{DBLP:journals/corr/abs-1812-02009} &   \multicolumn{2}{c|}{11} &   & \multicolumn{2}{c|}{11} &   & \multicolumn{2}{c|}{30} &   &   &  \bigstrut\\
    \hline \hline
    \cite{DBLP:journals/corr/abs-1904-03487} &   &   &   & \multicolumn{2}{c|}{22} & \multicolumn{1}{c|}{\footnotesize{\checkmark}
} & \multicolumn{2}{c|}{33} &   &   & \footnotesize{yes}
 \bigstrut\\
    \hline
    \cite{harz2018towards} &   &   &   &   &   &   &   & 10 & \footnotesize{\checkmark}
 &   &  \bigstrut\\
    \hline
    \cite{di2019survey} &   &   &   &   &   &   &   & 27 & \footnotesize{\checkmark}
 &   &  \bigstrut\\
    \hline
    \cite{DBLP:journals/corr/abs-1903-07602} &   &   &   &   &   &   &   &   &   & \multicolumn{1}{c|}{\footnotesize{yes}
} &  \bigstrut\\
    \hline
    \cite{DBLP:journals/corr/CachinV17,DBLP:journals/corr/abs-1711-03936,DBLP:journals/corr/abs-1805-02707,DBLP:journals/corr/abs-1904-04098} &   &   &   &   &   &   &   &   &   &   & \footnotesize{yes}
 \bigstrut\\
    \hline
    \end{tabular}%
  \label{table:related-workd}%
\end{table*}%

Table \ref{table:related-workd} highlights the relationship between the present work and related prior surveys, which accommodate some of vulnerabilities, attacks, defenses, and building-blocks. 
The most closely related survey is Atzei et al. \cite{atzei2017survey}, which discussed 12 types of vulnerabilities, 9 attacks, and 3 defenses in the context of Ethereum smart contracts (i.e. Blockchain 2.0). 
In contrast, we present a much more systematic treatment, by accommodating 44 types of vulnerabilities, 26 attacks, and 47 defenses.
From a methodological standpoint, we further discuss the root causes of vulnerabilities (e.g., the root cause of the {\em unchecked call return value} vulnerability is the {\em inconsistent exception handling} of Solidity). Moreover, we draw insights from the perspectives of vulnerability, attack, and defense.

There are a number of surveys on blockchain security and privacy which however have different perspectives than ours. 
First, Li et al. \cite{LI2017} reviewed security of blockchain technologies through the lens of 20 types of vulnerabilities, 6 attacks, and 5 defenses, without distinction between Blockchain 1.0 and 2.0 aspects. Similarly, Zhu et al. \cite{DBLP:journals/corr/abs-1812-02009} reviewed 11 smart contract vulnerabilities, 11 attacks against blockchain data, and 30 defenses. On the contrary, we focus on the Ethereum blockchain system by accommodating 44 types of vulnerabilities, 26 attacks, and 47 defenses. 

Second, Saad et al. \cite{DBLP:journals/corr/abs-1904-03487}
explored blockchains' attack surface in terms of cryptographic constructions, distributed system architecture, and applications; they cover 22 attacks and 33 defenses in Blockchain 1.0 and 2.0, but not vulnerabilities. 
Their survey is orthogonal to ours because (i) we focus on the Ethereum system, rather than multiple implementations of blockchains; (ii) we discuss vulnerabilities and their root causes as well as the attacks exploiting them, rather than attack surface; (iii) we provide insights into, and contrast, the vulnerabilities at different layers of the Ethereum architecture.

Third, Harz et al. \cite{harz2018towards} discussed 
10 smart contract verification tools. Similarly,
Angelo et al. \cite{di2019survey} discussed 27 tools for analyzing Ethereum smart contracts. On the contrary, we focus on the defenses of the Ethereum systems security, rather than purely on smart contracts. 

Fourth, Zhang et al. \cite{DBLP:journals/corr/abs-1903-07602} presented a comprehensive review on Bitcoin-like transactions and the underlying (cryptographic) mechanisms. Their review is geared towards the abstract blockchain model for Bitcoin-like transactions (i.e., Blockchain 1.0); in contrast, we focus on the Ethereum ecosystem, including the design and implementation of the blockchain platform and DApps.
There are surveys that focus on Bitcoin and cryptocurrencies \cite{7163021,8369416,7423672}, which use a transaction-centered model known as {\em unspent transaction outputs} (UTXO). In contrast, we focus on Ethereum, which uses an account-centered model.

Fifth, Orthogonal to the purpose of the present survey and tutorial, there are several surveys purely on blockchain consensus protocols \cite{DBLP:journals/corr/CachinV17,DBLP:journals/corr/abs-1711-03936,DBLP:journals/corr/abs-1805-02707,DBLP:journals/corr/abs-1904-04098}. In contrast, we focus on the overall Ethereum system.

\subsection{Paper outline}

Section \ref{sec:review-and-methodology} briefly reviews the Ethereum system and discusses the survey methodology.
Section \ref{sec:Vulnerability} presents the 44 Ethereum vulnerabilities according to the layers of the Ethereum architecture and a deeper analysis of their root causes.
Section \ref{sec:attacks} presents the 26 attacks against Ethereum and a deeper analysis on their consequences. 
Section \ref{sec:defenses} presents the 47 defenses, classified as proactive or reactive and a deeper analysis on their capabilities and investments. 
Section \ref{sec:further-discussion} discuss future research directions towards securing blockchain-based systems, including Ethereum as a special case.
Section \ref{sec:conclusion} conclude the present paper.

\section{Ethereum Review and Survey Methodology}
\label{sec:review-and-methodology}
\subsection{A brief review of the Ethereum system}
\begin{figure}[!htbp]
\centering
\includegraphics[width=.48\textwidth]{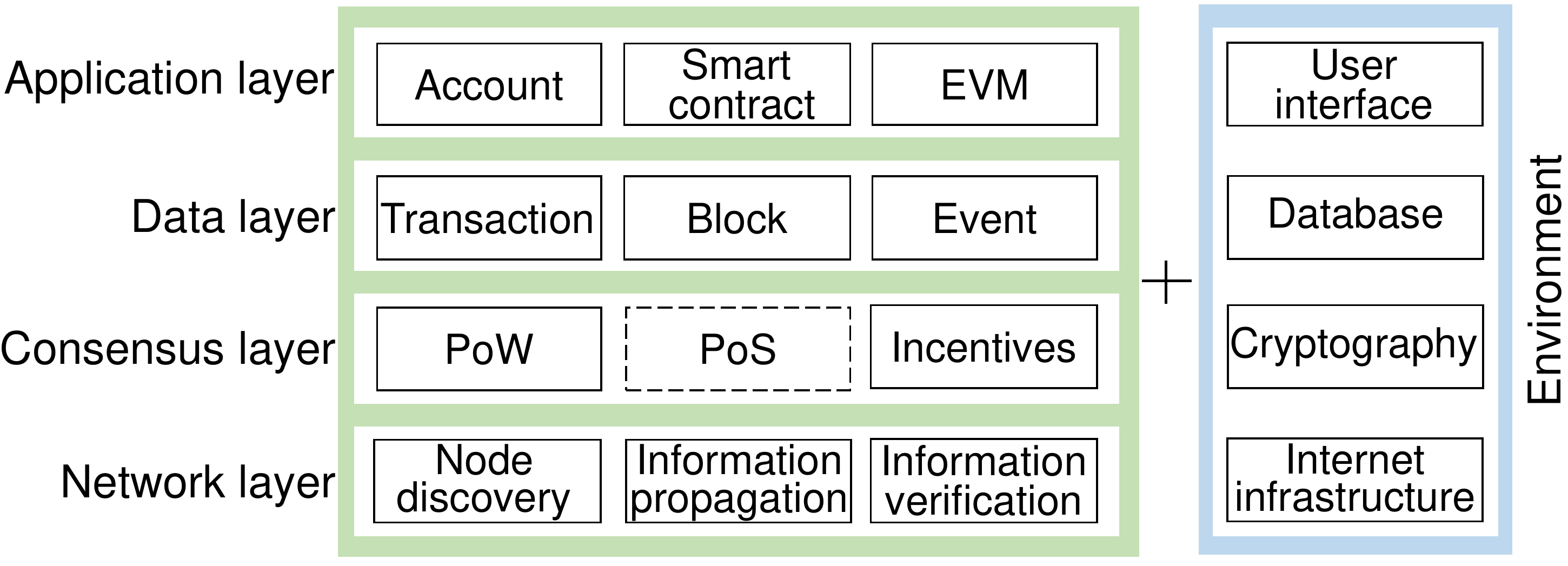}
\caption{Architecture of the Ethereum blockchain and its environment in which the Ethereum blockchain runs. 
\label{fig:ethereum-blockchain}}
\end{figure}

Figure \ref{fig:ethereum-blockchain} highlights a 4-layer architecture of the Ethereum blockchain, operates across these 4 layers.
At the {\it application layer}, Ethereum clients execute smart contracts, which are associated to Ethereum accounts, in EVM.
The {\it data layer} contains the blockchain data structures. The  {\it consensus layer} assures a consistent state of the blockchain. Note that Ethereum plans to replace its current use of Proof-of-Work (PoW) with a Proof-of-Stake (PoS).
The {\it network layer} formulates a Ethereum peer-to-peer (P2P) network of {\em nodes} or {\em clients} such that a node can always get the updated state of the blockchain from some of the active nodes. The {\it environment} serves these 4 layers via a corresponding component: a web user interface to interact with applications; databases for storing blockchain data; cryptographic mechanisms for supporting the consensus protocols; and Internet service for the network layer.

\begin{figure*}[!htbp]
\centering
\includegraphics[width=1.0\textwidth]{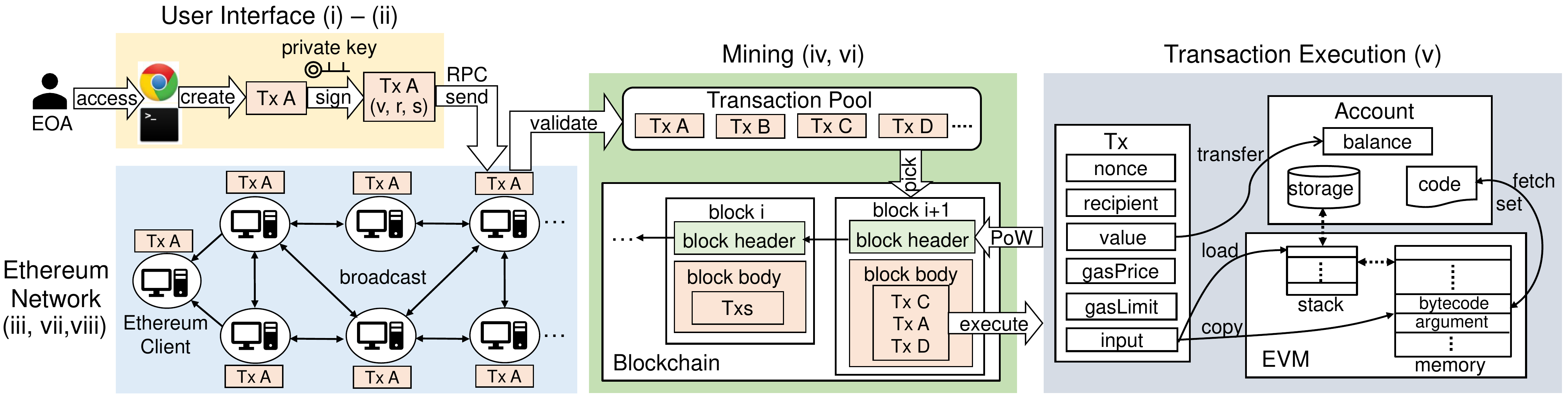}
\caption{An illustration of the lifecycle of an Ethereum transaction.
\label{fig:ethereum-workflow}}
\end{figure*}

\subsubsection{The application layer}
Ethereum supports two types of accounts: {\em externally owned accounts} (EOA) and {\em contract accounts}. 
An EOA is used to keep a user's funds in Wei, which is the smallest subdenomination of Ether and is worth $10^{-18}$ Ether.
An EOA is associated with, and addressed by, a public key; access to an EOA is authenticated by showing the ownership of the corresponding private key. In contrast, a contract account is associated with a piece of executable bytecode (i.e., smart contracts), which defines some business logic of interest. An EOA or contract account has a dynamic state, which is defined by: (i) $nonce$, which tracks the number of {\em transactions} that have been initiated by the owner of the EOA or the number of contracts created by the contract account; (ii) $balance$, which is the amount of Wei (i.e., $10^{-18}$ Ether) owned by the EOA or contract account; (iii) $storageRoot$, which is the hash of the root of the account's storage data structure {\em trie} that records a contract's state variables associated to the corresponding piece of bytecode (i.e., not applicable to EOA); 
(iv) $codeHash$, which is the hash value of a contract account's bytecode (i.e., not applicable to EOA). The state of a blockchain is defined by the states of the accounts on the blockchain.

Smart contracts are building-blocks of decentralized applications (DApps) running on top of the Ethereum blockchain. A DApp often has a user interface as its front-end and some smart contracts as its back-end. At the moment of writing, 2,497 DApps are running on top of Ethereum, including finance, governance, gambling, exchange, and wallet applications \cite{Stateoft17:online}. Some DApps issue their own cryptocurrency, called {\em tokens}, for purposes like Initial Coin Offering (ICO) and exchanges. A Ethereum-based token is a special kind of smart contract (e.g., ERC-20 \cite{ERC20Tok93:online}).

Smart contracts execute in EVMs, which are quasi-Turing-complete machines using a stack-based architecture; the term ``quasi'' means that the execution is limited by the amount of gas offered by the transaction in question \cite{omohundro2014cryptocurrencies}.

\subsubsection{The data layer}
A {\em transaction} is an interaction between an EOA (called {\em sender}) and anther EOA or contract account  (called {\em recipient}). A transaction is specified by: (i) $nonce$, which is a counter for tracking the total number of transactions that have been initiated by the sender; 
(ii) $recipient$, which specifies a transaction's destination EOA or contract account;
(iii) $value$, which is the amount of money (unit: Wei) to be transferred from the sender to the recipient (if applicable); 
(iv) $input$, which is the bytecode or data corresponding to the purpose of the transaction;
(v) $gasPrice$ and $gasLimit$, which respectively specify the unit price and the maximum amount of gas the sender is willing to pay the winning miner of a {\em block} containing the transaction;
(vi) $(v, r,s)$, which is the
Elliptic Curve Digital Signature Algorithm (ECDSA)
signature of the sender. 
The execution of a transaction updates the states of the accounts involved and therefore the state of the blockchain. 

Figure \ref{fig:ethereum-workflow} depicts  the lifecycle of a transaction. (i) A sender constructs a transaction and digitally signs it. (ii) The sender submits the signed transaction to an Ethereum client via a JSON-RPC call. 
(iii) The client validates the received transaction and broadcasts it to the Ethereum P2P network. 
(iv) Any client that receives the transaction and is a miner adds the transaction to its transaction pool. (v) A miner executes a sequence of transactions chosen from its transaction pool, formulates a new block, and updates the state of the blockchain as follows. For a {\tt money-transfer} transaction,
the specified $value$ is transferred from the sender's EOA to the recipient's EOA or contract account;
for a {\tt contract-creation} transaction where $input$ is a piece of bytecode, a new contract account is created and is associated with the bytecode; 
for a {\tt contract-invocation} transaction
where $recipient$ is a callee contract and $input$ uniquely identifies the callee function (and possibly some associated arguments), the bytecode associated to the callee contract account is loaded into the EVM. (vi) The miner solves a PoW by finding a random nonce such that the hash value of the block metadata in question is smaller than a certain threshold, which reflects the difficulty of creating a block. Unlike Bitcoin's computation-intense PoW \cite{nakamoto2008bitcoin}, Ethereum uses a memory-intense puzzle called ``Ethash'' \cite{wood2014ethereum}.
(vii) Upon creating a block, the miner broadcasts it to the Ethereum P2P network so that other clients can validate the block. (viii) Upon validating a block, the client in question appends the block to the blockchain.

\begin{figure}[!htbp]
\centering
\includegraphics[width=.45\textwidth]{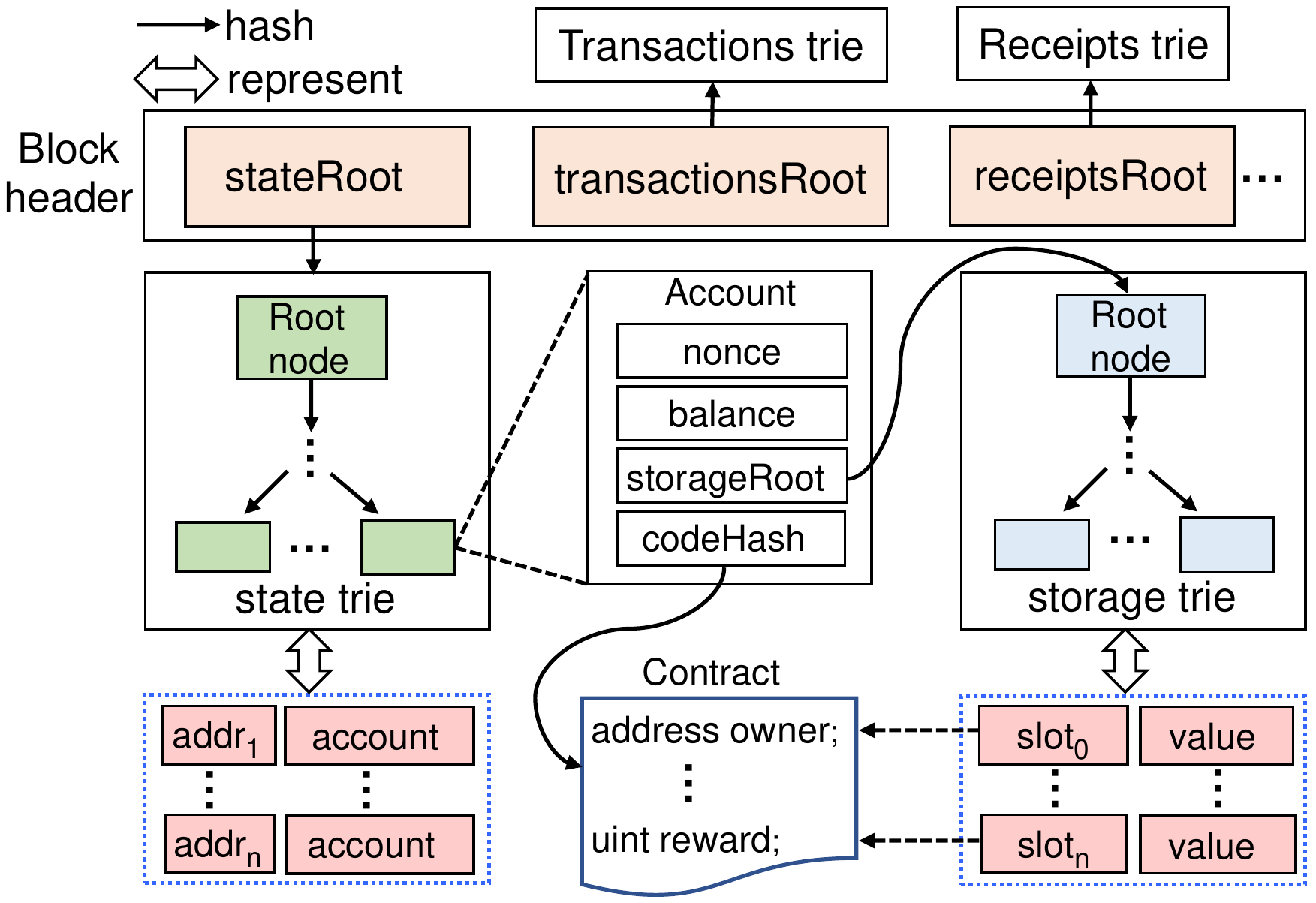}
\caption{Storage structure of Ethereum blockchain. 
\label{fig:data-storage}}
\end{figure}

Trie \cite{wood2014ethereum} is the data structure for storing Ethereum blockchain data (e.g., account states). Like a Patricia tree \cite{Patricia53:online}, a trie stores $(key, value)$ pairs while facilitating search as follows: the path from the root to a leaf node corresponds to a $key$ and the leaf node contains a $value$ (e.g., the state of an account). As illustrated in Figure \ref{fig:data-storage}, a block header may point to a state trie, a transaction trie (for bookkeeping transaction data), and a receipt trie (for bookkeeping the data related to the execution of transactions). Each contract account corresponding to a leaf or branch node on the state trie uses a separate storage trie to bookkeep the persistent data of the contract;
this storage trie also use a $(key,value)$ structure, where the position of each slot corresponds to a $key$ and the contract's state variable in each slot corresponds to a $value$. Note that an Ethereum blockchain has a {\em single} state trie because the state of the blockchain dynamically evolves.

\subsubsection{The consensus layer}
At the moment of writing, Ethereum takes about 14 seconds to create a block, meaning that multiple miners could create valid blocks simultaneously and that there could be many {\em stale} blocks. Ethereum uses a variant of the GHOST consensus protocol \cite{sompolinsky2015secure} to select the ``heaviest'' branch as the {\em main chain} where the ``heaviest'' branch is the sub-tree rooted at the fork in question and has the highest cumulative block difficulty \cite{garay2015bitcoin,wang2019survey}, while noting that {\em stale} blocks are not on the main chain.

\begin{figure}[!htbp]
\centering
\includegraphics[width=.48\textwidth]{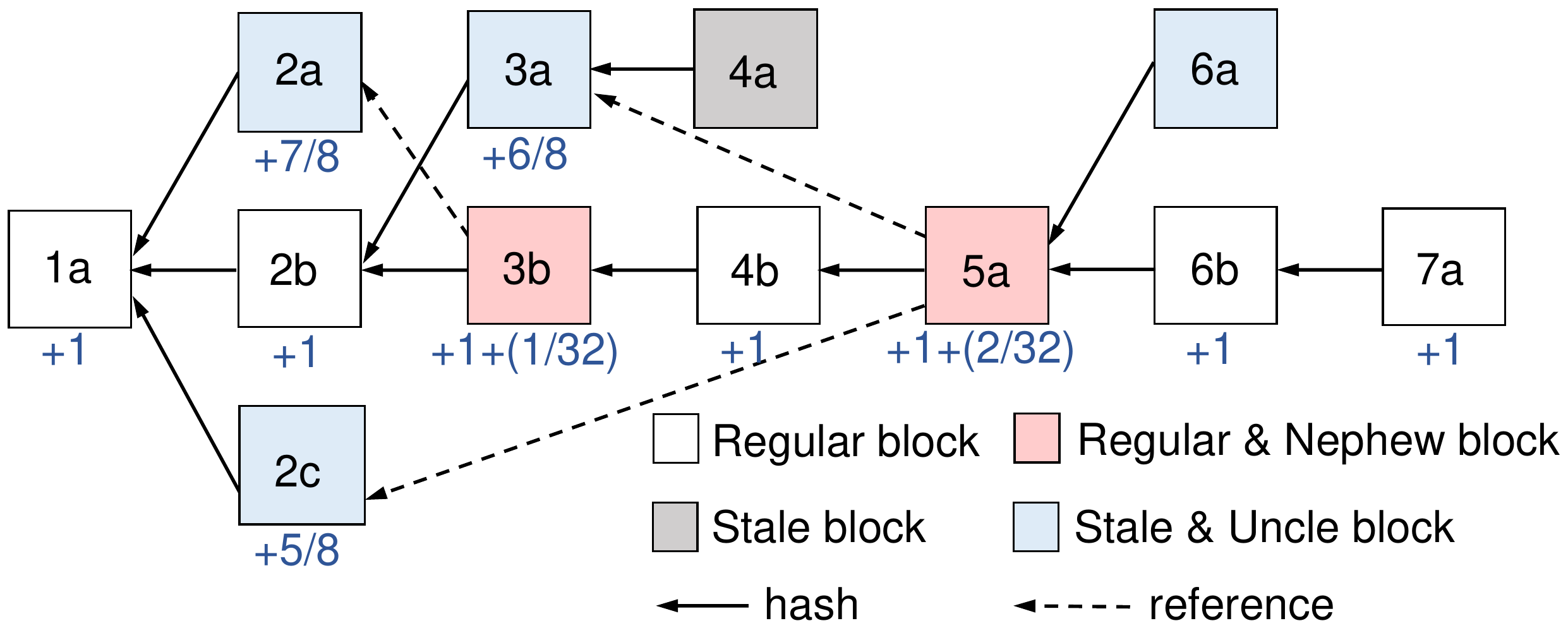}
\caption{A sub-tree of an Ethereum blockchain (adapted from \protect\cite{niu2019selfish}), where a value beneath a node is the total award to the miner of the node (unit: static block reward).
\label{fig:ethereum-reward}}
\end{figure}

Ethereum rewards not only the {\it regular} blocks on the main chain, but also the stale blocks referred by a regular block. As illustrated in Figure \ref{fig:ethereum-reward}, an {\em uncle} block is a stale block referenced by a regular block called {\em nephew} (via a dashed arrow). The distance between two blocks is their height difference on the tree. The miner of a regular block receives one unit of ``static block reward,'' which is worthy of 2 Ethers at the time of writing. In order to incentivize referencing to uncle blocks, the miner of a nephew regular block further receives $1/32$ of the static block reward for a reference (and for up to 2 references). The miner of the referenced uncle block is rewarded with $1-d/8$ of the static block reward, where $1\leq d\leq 6$ is the distance between the uncle block and the referencing nephew block. Miners of regular blocks also receive the gas fee specified by the transactions in question.

\subsubsection{The network layer}
The Ethereum network is a structured P2P network where each node (i.e., client) stores a copy of the entire blockchain.
For node discovery and routing purposes, 
each node maintains a dynamic routing table of 160 buckets and each bucket contains up to 16 entries of other nodes' IDs, IP address, UDP/TCP ports. Ethereum uses the RLPx protocol  \cite{kim2017measuring,marcus2018low} to discover target clients and uses the Ethereum Wire Protocol \cite{devp2pet75:online} to facilitate the exchange of Ethereum blockchain information (e.g., transactions, blocks) between clients.

\subsubsection{The environment}
The Ethereum blockchain runs in an environment, which naturally operates across the 4-layers to provide respective services, namely: a web interface for users to interact with the Ethereum blockchain; a database for Ethereum clients to store the blockchain data; cryptographic mechanisms for security purposes; and the Internet infrastructure to support blockchain networking and communication among Ethereum nodes.
We separate the Ethereum blockchain architecture from the environment because attacks against the Ethereum blockchain may come from the environment and these attacks may be better addressed in the environment rather than by the Ethereum blockchain, leading to a clean and modular abstraction.

\subsection{Survey methodology}
\subsubsection{Scope}
Since our focus is on Ethereum security, we will systematize the vulnerabilities of, the attacks against, and the defenses for Ethereum. Since these aspects are related to the programming language for writing smart contracts and the client software, we focus on the widely-used Solidity and Geth \cite{geth} as well as Parity \cite{parity-client}, respectively.

\subsubsection{Methodology}
Our methodology can be characterized as follows. First, we use a layered architecture to present the Ethereum ecosystem, from the application down to the data, the consensus, the network, and the environment layers. This layered view allows us to describe matters at the most appropriate layer.

Second, we consider security from three perspectives (i.e., {\em vulnerabilities}, {\em attacks}, and {\em defenses}) as well as the relationships between them. 
For each vulnerability, we discuss, among other things, its root cause and status (i.e., eliminated, can be eliminated by best practice, or still open). 
For each attack, we discuss, among other things, its cause, tactic, and direct impact.
For each defense, we discuss its mechanism and the (hidden) vulnerabilities it aims to protect from exploitation.
We aim to provide insights into the design and implementation of Ethereum as well as future research directions in order to address the range of open problems.

\section{Vulnerabilities}
\label{sec:Vulnerability}

\begin{figure*}[!htbp]
\centering
\includegraphics[width=.95\textwidth]{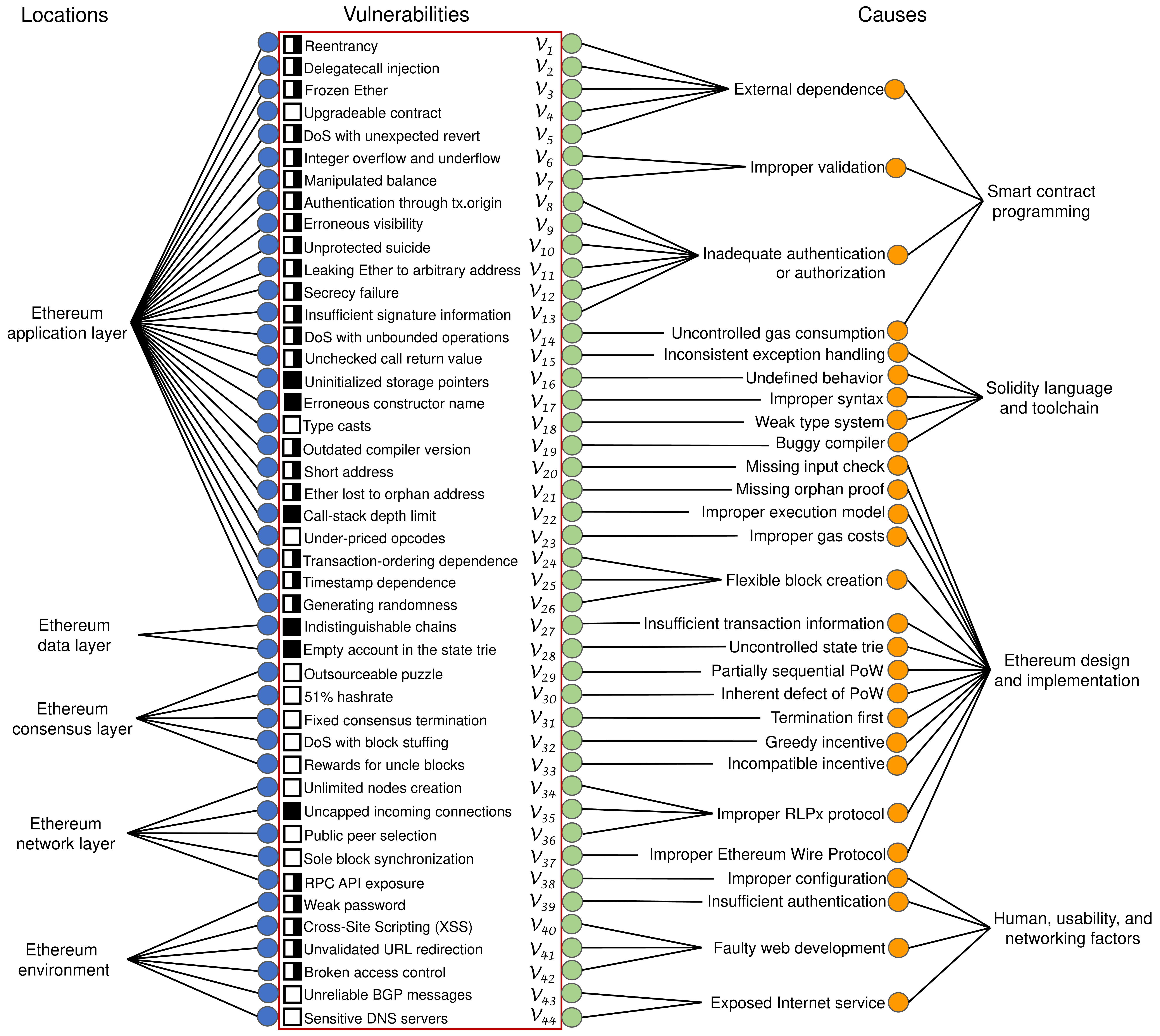}
\caption{A classification of Ethereum vulnerabilities and their state-of-the-art treatments ($\blacksquare$ means ``eliminated already'', $\squarerightblack$ means ``can be avoided by best practice'', and $\square$ means open (i.e., has yet to be eliminated).
\label{fig:vuls-cause-location}}
\end{figure*}

Figure \ref{fig:vuls-cause-location} highlights our classification of Ethereum vulnerabilities based on their {\em location}, {\em cause}, and {\em status} (i.e., {\em eliminated} vs. {\em can be avoided by best practice} vs. {\em open}). For ease of reference, we denote the 44 types of vulnerabilities as ${\cal V}_1,\ldots,{\cal V}_{44}$, respectively.
In what follows, we group them according to location. 

\subsection{Vulnerabilities at the application layer} 

\subsubsection{Reentrancy (${\cal V}_1$)} 
This vulnerability was first observed from the DAO attack \cite{DAOexploit}; its variants were later reported in \cite{rodler2018sereum}.
The vulnerability occurs when an external callee contract calls back to a function in the caller contract {\em before} the execution of the caller contract is completed (i.e., cyclic calls in a sense).
This allows the attacker to bypass the due validity check until the caller contract is drained of Ether or the transaction runs out of gas.
The vulnerability is caused by: (i) a contract's control-flow decision relies on some of its state variable(s) that should be, but are not, updated by the contract itself before calling another (i.e., an external) contract \cite{rodler2018sereum}; and (ii) there is no gas limit when handing the control-flow to another contract. The vulnerability can be prevented by one of the following methods \cite{Best_Practices}: (i) assuring that a contract's state variables are updated {\em before} calling another contract; (ii) introducing a {\myfont mutex} lock on the contract state to assure that only the lock owner can change the state; (iii) using the \textsf{transfer} method to send money to other contracts because this method only forwards 2,300 gas to the callee contract.

\subsubsection{Delegatecall injection  (${\cal V}_2$)} 
This vulnerability was first observed from an attack against the Parity wallet \cite{parity-hack}. In order to facilitate code-reuse, EVM provides an opcode, \textsc{delegatecall}, for embedding a callee contract’s bytecode into the bytecode of the caller contract as if it's a piece of the latter's bytecode \cite{krupp2018teether}. As a consequence, a malicious callee contract can directly modify (or manipulate) the state variables of the caller contract. This vulnerability is caused by the ability that the state variables of a caller contract can be updated by the bytecode of a callee contract. The vulnerability can be completely prevented by declaring a contract that is meant to be shared via the {\sf delegatecall} as a library, which is stateless \cite{Adrian_list}.

\subsubsection{Frozen Ether  (${\cal V}_3$)}
This vulnerability was first observed from another attack against the Parity wallet \cite{Security89:online}. The vulnerability results from the ability of users to deposit their money to their contract accounts with the inability to spend their money from those accounts, effectively freezing their money.
The vulnerability is caused by  \cite{jiang2018contractfuzzer}: (i) contracts not providing any function for spending money relying on the money-spending function of another contract (as a library) and (ii) the callee contract (i.e., the library) being killed accidentally or deliberately. The vulnerability can be prevented by assuring that mission-critical functions, or money-spending functions in this case, are not outsourced to another contract.

\subsubsection{Upgradable contract  (${\cal V}_4$)}
This vulnerability was first discussed in \cite{atzei2017survey}. The idea of contract upgrading was introduced to mitigate the problem that smart contracts, once deployed, cannot be modified even if they are later found to have vulnerabilities. In order to allow contract upgrading, there are two approaches: (i) splitting a contract into a {\em proxy contract} and a {\em logic contract} such that developers can upgrade the latter but not the former; and
(ii) using a {\em registry contract} to bookkeep the updated contracts. While effective, these approaches introduce a new vulnerability: when the contract developer is malicious, the updated (logic) contract can be malicious. The vulnerability is caused by one contract being able to call another contract, which can be malicious. This vulnerability (i.e., unsecure contact updating) remains to be an open problem.

\subsubsection{DoS with unexpected revert  (${\cal V}_5$)}
This vulnerability was first reported in \cite{Best_Practices}. It occurs either when a transaction is reverted due to a caller contract encountering a failure in an external call, or the callee contract deliberately performs the \textsf{revert} operation to disrupt the execution of the caller contract. 
This vulnerability is caused by the execution of a caller contract being reverted by a callee contract. This vulnerability can be prevented by letting a recipient invoke a transaction to ``pull'' the money that was set aside by a sender for the recipient, which effectively prevents a sender's transaction from being reverted \cite{Safety:online}. 

\subsubsection{Integer overflow and underflow  (${\cal V}_6$)} 
This vulnerability was first observed from the attack against the BEC tokens \cite{NVDCVE2014:online}. It occurs when
the result of an arithmetic operation falls outside of the range of a Solidity data type, causing (for example)  unauthorized manipulation to the attacker's balance \cite{NVDCVE2014:online} or other state variables. The vulnerability is caused by that Solidity source code does not perform proper validation on numeric inputs, and that neither the Solidity compiler nor the EVM enforces integer overflow/underflow detection.
This vulnerability can be prevented by using the
{\em SafeMath} library \cite{openzepp58:online} that handles these issues.

\subsubsection{Manipulated balance  (${\cal V}_7$)}
This vulnerability was first reported in \cite{HowtoSec70:online} and was also known as the “forcing Ether to contracts'' vulnerability.
This vulnerability occurs when a contract's control-flow decision relies on the value of {\myfont this.balance} or {\myfont address(this).balance}, which can be leveraged by an attacker to cause (for example) that only the attacker can obtain the money; see \cite{Adrian_list} for a detailed description. This vulnerability can be prevented by not using a contract’s balance in any condition check \cite{HowtoSec70:online}.

\subsubsection{Authentication through {\myfont tx.origin}  (${\cal V}_8$)} 
This vulnerability was first discussed in \cite{TxOrigin14:online}. The {\myfont tx.origin} is a global variable in Solidity and refers to the original EOA that initiates the transaction in question. This vulnerability occurs when a contract uses {\myfont tx.origin} for authorization, which can be compromised by a phishing attack. Figure \ref{fig:tx-origin} provides an example of this vulnerability (Line 6) in function {\sf withdrawAll}, which uses {\myfont tx.origin} to confirm that it is the owner of contract {\tt UserWallet} that is calling the function. However, if the owner of contract {\tt UserWallet} is tricked to transfer money to a malicious {\tt AttackContract}, the fallback function (Line 14) will be automatically executed, which in turn calls the {\sf withdrawAll()} function. Since the condition check (Line 6)
validates the EOA that initiated the transaction rather than the intermediate caller (i.e., {\tt AttackContract}),
the attacker can pass the authority check and steal the money in contract {\tt UserWallet}. This vulnerability can be prevented by using {\myfont msg.sender}, instead of {\myfont tx.origin}, for authentication \cite{Security10:online} because {\myfont msg.sender} returns the account that incurred the message.

\begin{figure}[!htbp]
\begin{lstlisting}[language=Solidity]
contract UserWallet {  // vulnerable contract
   address owner;
   constructor() public { owner = msg.sender; }
   ...
   function withdrawAll(address addr) public {
     require(tx.origin == owner);
     addr.transfer(this.balance);
   }
}

contract AttackContract { // malicious phishing contract
   address attacker; 
   function () payable {
     UserWallet(msg.sender).withdrawAll(attacker);
   }
}
\end{lstlisting}
\caption{The {\myfont tx.origin} vulnerability and its exploitations.}
\label{fig:tx-origin}
\end{figure}

\subsubsection{Erroneous visibility  (${\cal V}_9$)}
This vulnerability was first observed in an attack against the Parity wallet \cite{parity-hack}. It occurs when a function's visibility is incorrectly specified and thus permits unauthorized access. Specifically, Solidity provides four types of visibility to restrict access to a contract's functions, namely {\myfont public}, {\myfont external}, {\myfont internal}, and {\myfont private}, which respectively says a function can be called arbitrarily, only externally, only internally (i.e., within the contract and its derived contracts), or only within the contract. Functions that should not be called from an external contract should be specified as {\myfont private} or {\myfont internal}. However, Solidity makes functions as {\myfont public} by default, which allows attackers to directly call these improperly specified functions. Solidity (starting version 0.5.0) mitigates the vulnerability by making it mandatory for programmers to explicitly specify function visibility  \cite{Solidity53:online}. Still, this vulnerability cannot be completely prevented unless programmers correctly specify their functions' visibilities.

\subsubsection{Unprotected suicide  (${\cal V}_{10}$)}
This vulnerability was first observed from an attack against the Parity wallet \cite{Security89:online}. A contract account can be killed by the contract's owner (or a trusted third party) using the {\sf suicide} or {\sf self-destruct} method. A contract may have an ``owner'' whose privileges are typically specified when creating the contract. When a contract account is killed, its associated contract bytecode and storage are deleted forever. This vulnerability is caused by the  inadequate authentication enforced by a contract. The vulnerability can be mitigated by enforcing advanced authentication  (e.g, multi-factor authentication) to assure that a suicide operation must be approved by multiple parties \cite{SWC}.

\subsubsection{Leaking Ether to arbitrary address  (${\cal V}_{11}$)}
This vulnerability was first reported in \cite{nikolic2018finding}. The vulnerability occurs when a contract's funds can be withdrawn by any caller, who is neither the owner of the contract nor an investor who deposited funds to the contract. This vulnerability is caused by the failure in checking an caller’s identity when the caller invokes a function to send Ether to an arbitrary address. This vulnerability can be prevented by enforcing an adequate authentication on the functions for sending funds.

\subsubsection{Secrecy failure  (${\cal V}_{12}$)}
This vulnerability was first observed from a multi-player game in \cite{delmolino2016step} and was also called {\em keeping secrets} in \cite{atzei2017survey}. It can be exploited to benefit an attacker. In blockchain, restricting the visibility of a variable or function does not assure that the variable or function is secret, because of the public nature of blockchain (i.e., details of transactions are publicly known). Although restricting a state variable 
to be {\myfont private} can prevent other contracts from accessing it, any one can see the value of a state variable from the relevant transaction data. This vulnerability is caused by the lack of secrecy of sensitive data in an untrusted environment. A possible solution to preventing this vulnerability is to use cryptographic techniques, such as timed commitments \cite{delmolino2016step,atzei2017survey}. 

\subsubsection{Insufficient signature information  (${\cal V}_{13}$)}
This vulnerability was first exploited in a replay attack against smart contracts \cite{Replay-signatures}.
The vulnerability occurs when a digital signature 
turns out to be valid for multiple transactions, which can happen when one sender (say Alice) sends money to multiple recipients through a proxy contract (instead of initiating multiple transactions)
\cite{Replay-solidity}. In the proxy contract mechanism, Alice can send a digitally signed message off-chain (e.g. via email) to the recipients, similar to writing personal checks in the real world, to let the recipients withdraw money from the proxy contract via transactions. In order to assure that Alice does approve a certain payment, the proxy contract verifies the validity of the digital signature in question. However, if the signature does not give the due information (e.g., nonce, proxy contract address), a malicious recipient can replay the message multiple times to withdraw extra payments.
This vulnerability can be prevented by incorporating the due information in each message \cite{Programt93:online}.

\subsubsection{DoS with unbounded operations  (${\cal V}_{14}$)}
This vulnerability was first observed from the {\tt GovernMental} contract \cite{GovernMe0:online} and its variants were later discussed in \cite{KnownAtt55:online,grech2018madmax}. Recall that each block has a ``gas limit'' field that specifies the maximum total amount of gas that can be consumed by the transactions in a block. This vulnerability occurs when the amount of gas that is required for executing a contract exceeds the block gas limit. This vulnerability is caused by improper programming with unbounded operations in a contract (e.g., loop over a large array). This vulnerability can be mitigated by assuring (i) contracts should not use loops over data structures, especially those data structures that can be operated by EOA; and (ii) when a contract has to use loop over data structure, the contract should keep track of the loop and resume the aborted execution when the sender of the transaction re-invokes the same contract (in order to finish the execution of the contract) \cite{KnownAtt55:online}.

\subsubsection{Unchecked call return value  (${\cal V}_{15}$)}
This vulnerability was first discussed in \cite{luu2016making,SWC} and is also known as {\em mishandled exceptions}. 
It has two variants, called {\em gasless send} and {\em unchecked send} \cite{atzei2017survey,kalra2018zeus}.
Recall that Solidity provides two methods for a contract to call another: (i) directly referencing to a callee contract's instance; (ii) using one of the following four low-level methods: {\sf send}, {\sf call}, {\sf delegatecall} and {\sf callcode}.
There is a discrepancy in Solidity's handling of exceptions occurring in the execution of callee contracts  \cite{luu2016making}: if an exception occurs in case (i), the exception is automatically propagated back to the caller and the transaction is reverted entirely; if an exception occurs in case (ii), the callee contract returns {\sc false} back to the caller contract. This discrepancy can lead to unintended transactions unless the caller contract carefully addresses the discrepancy.
At the moment of writing, neither the Solidity compiler nor the EVM addresses the discrepancy. This vulnerability can be prevented by letting a caller contract check and address the discrepancy mentioned above.

\subsubsection{Uninitialized storage pointer  (${\cal V}_{16}$)}
This vulnerability was first reported in \cite{StorageA95:online}. Recall that in Solidity, the contract state variables are always laid out consecutively in storage, starting from slot $0$. For a complicated local variable (e.g., struct, array, or mapping), a reference is assigned to an unoccupied slot in the storage to point to the state variable. If the local variable is not explicitly initialized, then the local variable's reference points to slot $0$ by default, causing the content starting from slot $0$ to be overwritten \cite{Rob}. This vulnerability is caused by Solidity's treatment of uninitialized complicated local variables. This vulnerability has been eliminated by Solidity compiler, starting version 0.5.0, by reporting an error to contracts that contain uninitialized storage pointers \cite{Solidity53:online}.

\subsubsection{Erroneous constructor name  (${\cal V}_{17}$)}
This vulnerability was first observed from the {\tt Rubixi} contract \cite{Ethereum30:online}, where the constructor function has an incorrect name that allows anyone to become the owner of the contract. 
Prior to Solidity version 0.4.22, a function declared with the same name as the contract's is considered as the contract's constructor, which is only executed one-time upon the creation of the contract. If the constructor's name is misspelled by a programmer for whatever reason, the intended constructor becomes a public, normal function that can be invoked by {\em any} EOA. This vulnerability is caused by that Solidity does not provide a special syntax to distinguish a constructor function from a regular function \cite{construc42:online}. The vulnerability has been eliminated, starting Solidity version 0.4.22, by introducing the new keyword {\myfont constructor} \cite{ReleaseV79:online}.

\begin{figure}[!htbp]
\begin{lstlisting}[language=Solidity]
contract CounterLibrary { function add(uint) public returns (uint) }
contract CounterLib { function add(uint) public returns (uint) }
contract Game { 
   function play(CounterLibrary c) public { 
      c.add(1);
   } 
}
\end{lstlisting}
\vspace{-3mm}
\caption{The type casts vulnerability.}
\label{fig:type-cast}
\end{figure}

\subsubsection{Type casts  (${\cal V}_{18}$)}
This vulnerability was first reported in \cite{atzei2017survey}. Recall that a contract written in the Solidity language can call another contract by directly referencing to the callee contract's instance. As illustrated in Figure \ref{fig:type-cast}, contract {\tt Game} calls the function {\sf add()} in contract {\tt CounterLibrary} by referencing to its instance {\myfont c} (Line 5). When function {\sf play()} (Line 4) is invoked, the argument specifying the callee contract's address is cast to {\tt CounterLibrary}. However, the Solidity compiler can only check whether or not the {\tt CounterLibrary} contract declared function {\sf add()}, but cannot check whether or not the address argument conforms to that of the {\tt CounterLibrary} contract's. 
If the address associated to the {\tt CounterLib} contract (Line 2) contains a function that is named {\sf add()} and has the same declaration, then the {\sf add()} function in contract {\tt CounterLib} is executed, instead of the desired {\sf add()} function in contract {\tt CounterLibrary}. As a consequence, the EVM can be misled to run the attacker's contract. This vulnerability is caused by the incompetent type system of Solidity. Currently, there is no feasible way to avoid the vulnerability.

\subsubsection{Outdated compiler version  (${\cal V}_{19}$)} 
This vulnerability was first reported in \cite{SWC}. It occurs when a contract uses an outdated compiler, which contains bugs and thus makes a compiled contract vulnerable. This vulnerability can be prevented by using an up-to-date compiler.

\subsubsection{Short address  (${\cal V}_{20}$)}
This vulnerability was first discussed in \cite{HowtoFin93:online}. Recall that in a {\tt contract-invocation} transaction, the function selector and arguments are orderly encoded in the \textit{input} field, where the first four bytes specify the callee function and the rest data arranges arguments in chunks of 32 bytes. However, if the length of the encoded arguments is shorter than expected, EVM will auto-pad extra zeros to the arguments to make up for 32 bytes. Consider function {\sf transfer(address addr, uint tokens)} as an example. If the trailing one (i.e., last) byte of {\sf addr} is left off, two extra hex zeros will be added to the end of {\sf tokens}, which amplifies the number of tokens being sent.
This vulnerability is caused by that EVM does not check the validity of addresses \cite{Short-add}. This vulnerability can be prevented by checking the length of a transaction's input (i.e., {\myfont msg.data}) \cite{Worrysom16:online}.

\subsubsection{Ether lost to orphan address  (${\cal V}_{21}$)}
This vulnerability was first reported in \cite{atzei2017survey}. When transferring money, Ethereum only checks whether the length of the recipient's address is no greater than 160-bit but not the validity of the recipient's address. If money is sent to a non-existing orphan address, Ethereum automatically registers for the address than terminating the transaction. Since the address is not associated to {\em any} EOA or contract account, there is no way to withdraw the transferred money, which is effectively lost. This vulnerability is caused by that EVM is not orphan-proof. At the moment of writing, this vulnerability can only be prevented by manually assuring the correctness of the recipient's address.

\subsubsection{Call-stack depth limit  (${\cal V}_{22}$)}
This vulnerability was first reported in \cite{call-stack}. Recall that in the original specification of Ethereum execution model \cite{wood2014ethereum}, EVM's call-stack has a hard limit of 1024 frames. When a contract calls another contract, the call-stack depth of the transaction increases by one; when the number of nested calls exceeds 1024, Solidity throws an exception and aborts the call. An attacker can recursively call a contract, which may be deployed by the attacker, for 1023 times and then  calls a victim contract to reach the stack depth limit, which fails any subsequent external call made by the victim contract. Since Solidity does not propagate exceptions in low-level external calls, the victim contract may not be aware of the failure. 
This vulnerability is caused by EVM's inadequate execution model, and has been eliminated
by the hard fork for EIP-150, which re-defines the gas-consumption rules of external calls to make it impossible to reach 1024 in call stack depth \cite{EIP-150}.   

\subsubsection{Under-priced opcodes (${\cal V}_{23}$)}
This vulnerability was first observed from two DoS attacks \cite{spam-attack,Buterin,chen2017adaptive}. Recall that Ethereum uses the gas mechanism to prevent the abuse of computing resources (e.g., CPU, disk, network). This vulnerability occurs when a contract contains a lot of under-priced opcodes that consume a large amount of resources at a low gas cost, meaning that the execution of the contract wastes a lot of computing resources. This vulnerability is caused by the failure in properly setting the gas cost for consuming computing resources. In order to mitigate this vulnerability, Ethereum has raised the gas cost for the opcodes that were abused to launch the two DoS attacks described in \cite{EIP-150}. However, it is not clearly whether or not the vulnerability can be completely prevented by this mechanism or not \cite{chen2017adaptive}. 

\subsubsection{Transaction ordering dependence (a.k.a. front running;  ${\cal V}_{24}$)}
This vulnerability was first discussed in \cite{luu2016making}. It refers to the concurrency issue that the forthcoming state of blockchain depends on the execution order of transactions, which is however determined by the miners. Typically, miners group and order transactions into a new block based on the reward offered by the transactions. Since transactions are publicly broadcast to the network, a malicious EOA can offer a higher {\it gasPrice} to have its transactions assembled into blocks sooner than the others'. Moreover, a malicious miner can always pick up its own transactions regardless of the {\em gasPrice}. This vulnerability is caused by that the state of a contract depends on how minors select transactions to assemble blocks. This vulnerability can be mitigated by using a cryptographic commit-reveal scheme to hide the information (e.g., {\em gasPrice}, {\em value}) offered by  transactions \cite{security25:online, ToSinkFr74:online}, or by introducing a guard condition to assure that an invocation of a contract either
returns the expected output or fails \cite{luu2016making}.

\subsubsection{Timestamp dependence (${\cal V}_{25}$)}
This vulnerability was first reported in \cite{luu2016making}. It occurs when a contract uses the {\myfont block.timestamp} as a part of the triggering condition for executing a critical operation (e.g., money transfer) or as the source of randomness, which however can be manipulated by a malicious miner. The vulnerability is caused by that Ethereum only prescribes that a timestamp must be greater than the timestamp of its parent block and be within future 900 seconds of the current clock. If a contract uses a timestamp-based condition (e.g., {\myfont block.timestamp} $\% 25 == 0$) to determine whether or not to transfer money, a malicious miner can slightly shift the timestamp to satisfy the condition to benefit the attacker. This vulnerability can be prevented by not using {\myfont block.timestamp} in any decision-making conditions.

\subsubsection{Generating randomness (${\cal V}_{26}$)}
This vulnerability was first reported in \cite{Predicti72:online}. Many gambling and lottery contracts select winners randomly, for which a common practice is to generate a pseudorandom number based on some initial private seed (e.g., {\myfont block.number}, {\myfont block.timestamp}, {\myfont block.difficulty} or {\myfont blockhash}). However, these seeds are fully controlled by miners, meaning that a malicious miner can manipulate these variables to make itself the winner \cite{randomness,fravoll}.
This vulnerability is caused by manipulable entropy sources. There are several proposals for addressing this problem, and each proposal has its own pros and cons. For example, the Oracle RNG proposal \cite{oracle} uses existing external services to generate random numbers off-chain and then send back to the requesting contract, meaning that there is a single-point-of-failure in the Oracle RNG; the RANDAO proposal \cite{RANDAO} uses a collaborative cryptographic commit-reveal scheme to generate a random number by multiple participants altogether, meaning that it achieves a limited throughput.

\subsection{Vulnerabilities at the data layer} 

\subsubsection{Indistinguishable chains (${\cal V}_{27}$)}
This vulnerability was first observed from the cross-chain replay attack when Ethereum was divided into two chains, namely ETH and ETC \cite{Replay-coindesk}. Recall that  Ethereum uses ECDSA to sign transactions. 
Prior to the hard fork for EIP-155 \cite{EIPseip146:online}, each transaction consisted of six fields (i.e., \textit{nonce}, \textit{recipient}, \textit{value}, \textit{input}, \textit{gasPrice} and \textit{gasLimit}). However, the digital signatures were not chain-specific because no chain-specific information was even known back then. 
As a consequence, a transaction created for one chain can be reused for another chain. This vulnerability has been eliminated by incorporating and signing a chainID.

\subsubsection{Empty account in the state trie (${\cal V}_{28}$)}
This vulnerability was first observed from a DoS attack reported in \cite{Buterin,chen2017adaptive}. An {\em empty} account is an account that has zero nonce, zero balance, no code associated to it, and no storage associated to it. An empty account is functionally equivalent to a non-existing account, except that an empty account needs to be bookkept in the Ethereum state trie and thus increases the synchronization and transaction processing time. This means that an attacker can incur a large number of empty accounts to substantially increase the the synchronization and transaction processing time, effectively causing a DoS attack \cite{Buterin,chen2017adaptive}.   
An empty account can be incurred by an attacker using the {\sc suicide} opcode to transfer zero Ethers to a non-existing account. 
This vulnerability was caused by the lack of control over that an empty account should not be included in the state trie.
This vulnerability has been eliminated by the hard fork for EIP-161 \cite{EIPseip189:online}, which removes those empty accounts from the state trie and prevents any empty account from being stored in the state trie.

\subsection{Vulnerabilities at the consensus layer} 

\subsubsection{Outsourceable puzzle (${\cal V}_{29}$)}
This vulnerability was reported in \cite{wang2019survey}. Recall that Ethereum adopts the PoW puzzle called Ethash, which was meant to be ASIC-resistant and be able to limit the use of parallel computing (owing to
that the vast majority of a miner's effort will be reading a dataset via the limited memory bandwidth). However, a crafty miner can still divide the task of searching for a puzzle solution into multiple smaller tasks and then outsource them. This vulnerability is caused by that Ethash only makes the puzzle solution partially sequential in preimage search, rather than relying on an inherently sequential PoW. Several puzzles are proposed to cope with this problem \cite{miller2015nonoutsourceable,daian2017short}, which however have not been adopted by the Ethereum community.

\subsubsection{51\% hashrate (${\cal V}_{30}$)}
This vulnerability is inherent to PoW-based blockchains, where an attacker controlling a majority of the mining power can take over the blockchain \cite{li2017survey,saad2019exploring}. Such an attacker can reverse transactions by formulating blocks and performing double-spending at will. This vulnerability is inevitable to PoW-based consensus protocols.

\subsubsection{Fixed consensus termination (${\cal V}_{31}$)} 
This vulnerability was first discussed in \cite{gramoli2017blockchain}. It refers to the fact that Ethereum's consensus protocol achieves a probabilistic agreement with a deterministic termination, meaning that a block is considered persistent in the main chain with a high probability if it is followed by a fixed number of $m$ blocks. In other words, the consensus for block $i$ will terminate when the chain depth reaches $i+m$ and all of the transactions contained in block $i$ are committed (i.e., a merchant can take an external action, such as shipping goods). However, the probability of agreement can be affected by various factors \cite{gramoli2017blockchain}. For example, in the presence of communication delay, applications should wait for more blocks to be confirmed so as to achieve a higher security against double-spending. This vulnerability is caused by that no deterministic protocol can simultaneously guarantee {\em agreement}, {\em termination}, and {\em validity} in an asynchronous network \cite{fischer1982impossibility}. 

\subsubsection{DoS with block stuffing (${\cal V}_{32}$)}
This vulnerability was first observed from the Fomo3D contract \cite{Fomo3D-4}. The vulnerability is that during a period of time, only the attacker's transactions are included in the newly-mined blocks and the other transactions are abandoned by the miners. This can happen when the attacker offers a higher {\it gasPrice} to incentivize the miners to select the attacker's transactions. This vulnerability is caused by the greedy mining incentive mechanism. At the moment of writing, there is no solution to prevent this vulnerability. 

\subsubsection{Rewards for uncle blocks (${\cal V}_{33}$)}
This vulnerability was independently reported in \cite{gervais2016security,niu2019selfish,ritz2018impact}. The vulnerability refers to that the uncle-rewarding mechanism reviewed above can lower the security of Ethereum when compared with Bitcoin. In particular, the uncle-rewarding mechanism reduces the risk for selfish miners that selectively release blocks to maximize their own profit because stale blocks may become uncle blocks and then receive rewards; this essentially incentivize selfish mining and double-spending. At the moment of writing, it is not clear how to eliminate this vulnerability.

\subsection{Vulnerabilities at the network layer} 
 
\subsubsection{Unlimited nodes creation (${\cal V}_{34}$)}
This vulnerability was reported for the Geth client prior to its version 1.8 \cite{marcus2018low}. In the Ethereum network, each node is identified by a unique ID, which is a 64-byte ECDSA public key. An attacker could create an unlimited number of nodes on a single machine (i.e., with the same IP address) and use these nodes to monopolize the incoming and outgoing connections of some victim nodes, effectively isolating the victims from the other peers in the network. This vulnerability is caused by the weak restriction on the node generation process.
This vulnerability can be eliminated by using a combination of IP address and public key as node ID. This countermeasure has not been adopted by the Geth developers who argue that it has a negative impact on the usability of the client.

\subsubsection{Uncapped incoming connections (${\cal V}_{35}$)}
This vulnerability was in the Geth client prior to its version 1.8 \cite{marcus2018low}. Each node can have a total number of {\myfont maxpeers} (with a default value 25) connections at any point in time, and can initiate up to $\lfloor(1 + \text{\myfont maxpeers})/2\rfloor$ outgoing TCP connections with the other nodes.
However, there was no upper limit on the number of incoming TCP connections initiated by the other nodes. This gives the attacker an opportunity to eclipse a victim by establishing an {\myfont maxpeers} many of incoming connections to a victim node, which has no outgoing connections. This vulnerability has been eliminated in Geth v1.8 by enforcing an upper limit on the number of incoming TCP connections to a node, with a default value $\lfloor\text{\myfont maxpeers}/3\rfloor = 8$.

\subsubsection{Public peer selection (${\cal V}_{36}$)}
This vulnerability was detected in Geth client prior to its version 1.8 \cite{marcus2018low}. Recall that the Ethereum P2P network uses a modified Kademlia DHT \cite{maymounkov2002kademlia} for node discovery and that each node maintains a routing table of 256 {\it buckets} for storing information about the other nodes. The buckets are arranged based on the XOR distance between a node's ID and its neighboring node's ID \cite{kim2018measuring}. When a node, say $A$, needs to locate a target node, $A$ queries the 16 nodes in its bucket that are relatively close to the target node and asks each of these 16 nodes, say $B$, to return the 16 IDs of $B$'s neighbors that are closer to the target node. The process iterates until the target node is identified. However, the mapping from node IDs to buckets in the routing table is public, meaning that the attacker can freely craft node IDs that can land in a victim node's buckets and insert malicious node IDs into the victim node’s routing table \cite{marcus2018low}. This vulnerability can be limited by making
the ``node IDs to buckets'' mapping private. This countermeasure has not been adopted by the Geth developers who argue that it has a negative impact on the usability of the client.

\subsubsection{Sole block synchronization (${\cal V}_{37}$)}
This vulnerability was first reported in \cite{wust2016ethereum}. It allows an attacker to partition the Ethereum P2P network without monopolizing the connections of a victim client. Recall that each block header contains a {\em difficulty} field, which records the mining difficulty of the block. The total difficulty of the blockchain, denoted by {\em totalDifficulty}, is the sum of the difficulty of the blocks up to the present one. When a client, say $A$, receives from, say client $B$, a block of which the difference {\em totalDifficulty} $ - $ {\em difficulty} is greater than the {\em totalDifficulty} at the blockchain stored on client $A$ (meaning that client $A$ missed a number of blocks), $A$ should start a block synchronization with $B$. Ethereum only allows a client to synchronize with one other client at a time (for network load considerations). This means that if client $B$ is malicious and deliberately delays the synchronization in response to $A$'s request, the blockchain at client $A$ is stalled and $A$ rejects every subsequent block, which may facilitate double-spending and DoS attacks. This vulnerability can be mitigated by synchronizing $A$ with multiple nodes, which however increases the network load.

\subsubsection{RPC API exposure (${\cal V}_{38}$)}
This vulnerability was first observed from the attack against the Geth and Parity clients \cite{Security_RPC}. The JSON-RPC of Ethereum clients provide various APIs for EOAs to communicate with the Ethereum network. For security purposes, the interface should only open locally and not be accessible from the internet. However, the standard port 8545 assigned to JSON-RPC can be accessed remotely in the Geth and Parity clients by default, which makes it possible for an attacker to call these remote clients via a JSON request \cite{wang2018attack}. Once having access to the remote client, the attacker can obtain sensitive data and perform certain unauthorized actions on the remote client. The vulnerability is caused by insecure API design and improper configuration. The vulnerability can be prevented by configuring the listening port (rather than using the default one) and adding access control to filter remote RPC calls.

\subsection{Vulnerabilities in the Ethereum environment} 
We propose considering vulnerabilities in the environment with which Ethereum interacts because they also pose as threats to Ethereum.

\subsubsection{Weak password (${\cal V}_{39}$)} 
This vulnerability was first observed from an attack against the DApp called Enigma \cite{Enigma}.
The vulnerability refers to not only low-entropy passwords but also password reuse and insecure password storage. When the password of a DApp administrator is compromised, the attacker can manipulate the DApp's webpage at will to attack others.

\subsubsection{Cross-Site Scripting (XSS;  ${\cal V}_{40}$)} 
This vulnerability was first observed from the attack against a cryptocurrency exchange, EtherDelta \cite{EtherDelta}.
XSS is a browser-side vulnerability which allows an attacker to inject malicious JavaScript code into a HTTP webpage. To initiate a transaction via web-based DApp, an EOA may need to fetch its private key from a local file to the browser. An attacker can steal the private key by uploading a malicious JavaScript to the webserver, while claiming (for example) that the attacker is publishing information about exchanging some tokens. The webserver pushes the attacker's token-exchange content to a victim's browser, which executes the malicious JavaScript code, causing the private key to be exposed to the attacker. This vulnerability can be prevented by letting a user sign transactions offline and then initiate  transaction online, meaning that the private key is never loaded into the browser and cannot be compromised by the attack mentioned above.

\subsubsection{Unvalidated URL redirection (${\cal V}_{41}$)} 
This vulnerability was first observed from an attack against the Ethereum wallet called MyEtherWallet \cite{MyEtherWallet}.
Web applications frequently redirect users to other webpages. If the target URL is not authenticated, an attacker can replace the target URL with a phishing or malicious website, which renders the users vulnerable to attacks.

\subsubsection{Broken access control (${\cal V}_{42}$)} 
This vulnerability was first observed from an attack against the decentralized application called CoinDash \cite{CoinDash}. The access control failure at the server side causes an unauthorized access to the web application, which may result in the manipulation of the web page. Many DApps start with an Initial Coin Offering (ICO). A common practice is that participants obtain tokens by depositing Ether to a certain contract account, whose address is published on the website of the DApp. If an attacker breaks into the webserver and replaces the original contract address with its own, then the coin buyers will pay the attacker.

\subsubsection{Unreliable Border Gateway Protocol (BGP) messages (${\cal V}_{43}$)} 
This vulnerability was first observed from an attack against MyEtherWallet \cite{MyEtherWallet}.
Each BGP router maintains a routing table with routing paths between ASes. BGP peers broadcast IP prefixes they own and dynamically update their routing tables according to the IP prefixes they received from their neighbors. 
However, BGP has no mechanism for validating messages communicated between ASes, and therefore simply treats them as trusted \cite{mitseva2018state}. An attacker can announce ownership of any prefix to its neighbor routers to redirect traffic towards the malicious ASes under its control. Then, the attacker can pretend to be a DNS server and redirect victims to malicious websites at IP addresses under the attacker's control.

\subsubsection{Sensitive Domain Name System (DNS) servers  (${\cal V}_{44}$)} 
This vulnerability was also observed from the attack against EtherDelta \cite{Cryptocu21:online}. A wide range of vulnerabilities that can be exploited to incorrectly resolve DNS queries and redirect users to malicious websites, which can be achieved by hijacking DNS through cache poisoning, stale records, and compromising the resolver configuration file. \cite{kang2016domain}.

\subsection{Further analysis of vulnerability causes} 

Now we present a taxonomy of the root causes of the vulnerabilities reviewed above. As highlighted in Figure
\ref{fig:vuls-cause-location}, the vulnerabilities are caused by incompetence or flaws in {\em smart contract programming}, {\em solidity language and toolchain}, {\em Ethereum design and implementation}, and {\em human, usability, and networking factors}.

\subsubsection{Smart contract programming}
These causes can be further divided into four sub-causes: {\em external dependence}, meaning a contract's execution relies on the behavior of an external contract; {\em improper validation}, meaning a failure in checking a condition allows the passing of an invalid input; {\em inadequate authentication or authorization}, causing failures in checking a caller's identity or privilege when the caller attempts to access a protected data item or functionality; and {\em uncontrolled gas consumption}, meaning a failure in gas allocation permits a DoS attack. These causes led to 14 types of vulnerabilities. Among these 14,
only 4 (i.e., {\em Integer overflow and underflow (${\cal V}_{6}$)}, {\em erroneous visibility} (${\cal V}_{9}$), {\em secrecy failure} (${\cal V}_{12}$), {\em DoS with unbounded operations} (${\cal V}_{14}$)) exist in traditional software, meaning that the other 10 are unique to Ethereum smart contract programming. This leads to:

\begin{insight}
\label{insight:ethereum-introduce-new-vulnerabilities}
Ethereum smart contracts introduce new kinds of vulnerabilities.
\end{insight}

Insight \ref{insight:ethereum-introduce-new-vulnerabilities} is interesting because it says that a new programming paradigm  demands adequate training for programmers' secure coding practice. Indeed, among the 26 types of application-layer vulnerabilities, 20 can be prevented by using best practices, highlighting the importance of engineering best practices.

\subsubsection{Solidity language and tool chain}
These causes can be further divided into five sub-causes, among which four sub-causes are related to the improper design of the Solidity language (i.e., {\em inconsistent exception handling} between direct call and low-level calls; {\em undefined behavior} of uninitialized storage pointers; {\em improper syntax} of constructor function; and {\em weak type system} with flexible typing rules) and the other sub-cause is {\em buggy compiler} (i.e., insufficient tool chain support). These five causes contributed 5 types of vulnerabilities. This leads to the following insight, which is clearly interesting and important.

\begin{insight}
It is important to design more secure programming languages and have more secure supporting tools for programmers to write secure smart contracts. 
\end{insight}

\subsubsection{Ethereum design and implementation}
These causes can be further divided into 14 sub-causes, which are related to, among other things, EVM, blockchain, PoW consensus, incentive mechanism, and P2P protocol. These causes cut across the application layer, the data layer, the consensus layer, and the network layer. The root causes related to EVM include: (i) {\em missing input check}, meaning no check on the validity of a transaction's data; (ii) {\em missing orphan proof}, meaning no check on a nonexistent recipient address; (iii) {\em improper execution model}, meaning the behavior of EVM is not properly specified; and (iv) {\em improper gas costs}, meaning the gas costs of EVM opcodes are not properly specified. 

The root causes related to blockchain include: (i) {\em flexible block creation}, meaning there are no restrictions on miners when they creating blocks, allowing them to create blocks to favor themselves; (ii) {\em insufficient transaction information}, meaning that a transaction can be accepted by multiple Ethereum blockchains (e.g, ETH, ETC, Ropsten), rather than a specific blockchain, owing to the lack of information specifying a unique blockchain; and (iii) {\em uncontrolled state trie}, meaning that there are no restrictions on the accounts that can be stored in the state trie. 

The root causes related to the PoW consensus protocol include: 
(i) {\em partially sequential PoW}, meaning that Ethereum's PoW puzzle can still be outsourced to different miners; (ii) {\em inherent defect of PoW}, meaning that a miner with 51\% hashrate can extend the main chain at will; and (iii) {\em termination first}, meaning that the Ethereum consensus protocol prefers termination to agreement. 

The root causes related to incentive mechanism include: (i) {\em greedy incentive}, meaning that a miner always selects transactions with higher $gasPrice$; and (ii) {\em incompatible incentive}, meaning that a miner can get a higher payoff by deviating from the consensus protocol. 

The root causes related to the P2P protocol include: (i) {\em improper RLPx protocol}, meaning that the node discovery and routing algorithms are not properly designed; and (ii) {\em improper Ethereum wire protocol}, meaning that the blockchain synchronization algorithm is not properly designed.

The root causes mentioned above contributed 18 vulnerabilities, cutting across the application layer, the data layer, the consensus layer, and the network layer; 9 out of these 18 vulnerabilities are still largely open (i.e., having yet to be eliminated). In summary, we draw:

\begin{insight}
\label{insight:blockchain-is-hard-to-design}
The vulnerabilities caused by the design and implementation of the Ethereum blockchain are harder to cope with than the other vulnerabilities.
\end{insight}

Insight \ref{insight:blockchain-is-hard-to-design} is interesting because it highlights the difficulty in designing and implementing a secure blockchain platform, despite (as we will see later) that the vulnerabilities in blockchain may not cause as big losses as the vulnerabilities at the application layer.  

\subsubsection{Human, usability and networking factors}
These causes can be further divided into four sub-causes: {\em improper configuration}, meaning that an Ethereum client is installed with incorrect permissions; {\em insufficient authentication}, meaning the use of weak or exposed passwords; {\em faulty web development}, meaning that a DApp's web interface is vulnerable; and {\em exposed Internet service}, meaning attacks coming from the Internet are not blocked.
These causes led to seven vulnerabilities.
In summary, we draw the following interesting insight.

\begin{insight}
Vulnerabilities in the Ethereum environment are largely caused by human, usability, and networking factors, which however consist of a smaller fraction of the vulnerabilities when compared with the vulnerabilities that are inherent to the Ethereum blockchain and smart contracts.
\end{insight}

\subsubsection{Cross-cutting analysis}
We observe that there are 13 vulnerabilities that are largely open. In order to draw insights into the approaches that may be able to eliminate them, we look deeper into their root causes (than what were mentioned above). 
First, we consider the vulnerabilities that are inherent to the Ethereum blockchain design, including: {\em under-priced opcodes} (${\cal V}_{23}$),  which are inherent to the EVM gas cost mechanism; {\em outsourceable puzzle} (${\cal V}_{29}$), {\em 51\% hashrate} (${\cal V}_{30}$), and {\em fixed consensus termination} (${\cal V}_{31}$), which are inherent to the PoW consensus mechanism; and {\em DoS with block stuffing} (${\cal V}_{32}$) and {\em rewards for uncle blocks} (${\cal V}_{33}$), which are inherent to the Ethereum incentive mechanism. In order to eliminate or mitigate these vulnerabilities, alternate gas cost mechanisms, consensus mechanisms, and incentive mechanisms may have to be used. 

Second, we consider the vulnerabilities that are inherent to the implementation of Ethereum clients, including
{\em unlimited nodes creation} (${\cal V}_{34}$), {\em public peer selection} (${\cal V}_{36}$), and {\em sole block synchronization} (${\cal V}_{37}$). In order to eliminate or mitigate these vulnerabilities, alternate network protocols may have to be used without sacrificing platform usability or network performance.

Third, we consider the vulnerabilities that are inherent to the insecure environment in which Ethereum runs, including {\em unreliable BGP messages} (${\cal V}_{43}$) and {\em sensitive DNS servers} (${\cal V}_{44}$). In order to eliminate or mitigate these vulnerabilities, extra effort must be made to defend the environment regardless how Ethereum is designed and implemented. 

Fourth, we consider the vulnerabilities that are inherent to the application-layer flexibility, including {\em upgradable contract} (${\cal V}_{4}$) and {\em type casts} (${\cal V}_{18}$). In order to eliminate or mitigate these vulnerabilities, alternate contract design pattern and type system may have to be used.

Summarizing the preceding discussion, we draw:

\begin{insight}
There are many vulnerabilities that are yet to be tackled in order to adequately defend Ethereum (or blockchain-based DApp systems in general), cutting across of every layer of the architecture, from the application layer, the data layer, the consensus layer, the network layer, down to the environment layer.
\end{insight}

\section{Attacks}
\label{sec:attacks}

Corresponding to the presentation of vulnerabilities, we group the 26 attacks we consider according to the locations of the vulnerabilities they exploit.
We describe each attack from the following perspective: its {\em history}, {\em cause}, {\em tactic}, and {\em direct impact} (i.e., the consequence when the attack succeeds). For ease of reference, we denote the 26 attacks by  ${\cal A}_1,\ldots,{\cal A}_{26}$, respectively. Note that some of these 26 attacks may correspond to the same type of attacks (i.e., sharing the same name), but they exploit different vulnerabilities or vulnerability combinations and/or cause different consequences. For example, both $\A_2$ and $\A_3$ correspond to the {\em parity multisignature wallet attack}; however, $\A_2$ exploits the {\em delegatecall injection} vulnerability ($\V_2$) and the {\em erroneous visibility} vulnerability ($\V_9$) to cause the consequence of {\em unauthorized code execution}, while $\A_3$ exploits the {\em frozen ether} vulnerability ($\V_3$) and the {\em unprotected suicide} vulnerability ($\V_{10}$) to cause the consequence of {\em DoS}.

\subsection{Attacks at the application layer}

\subsubsection{The DAO attack (${\cal A}_1$)} 
\begin{figure}[!htbp]
\includegraphics[width=.46\textwidth]{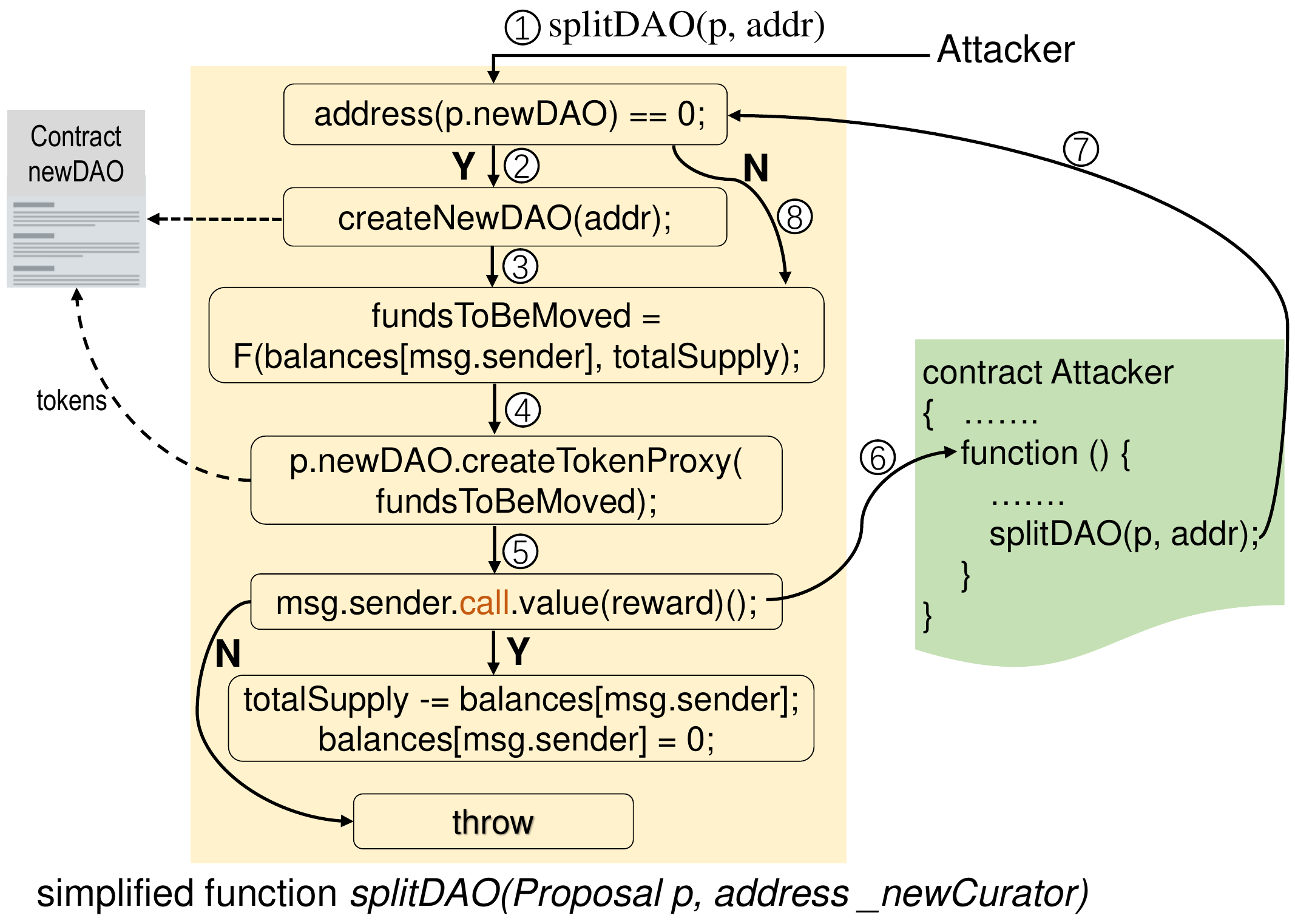}
\centering
\caption{The DAO attack exploiting the reentrancy vulnerability ($\V_1$).
\label{fig:TheDAO}}
\end{figure}

The contract {\tt DAO} is a financial application running on top of Ethereum. In June 2016, it was attacked to cause the loss of US\$60M \cite{TheDAO}. The financial application is that
investors (i.e., token holders) vote on investment proposals for spending their money (i.e., ``investment crowdsourcing''). 
Once a proposal is approved by a majority, the amount of money approved by the supporting investors is moved to the proposer's account, and the amount of money owned by the minority that opposes the proposal is respectively ``refunded'' to each of them via some newly created contract accounts. This mechanism was implemented in the {\sf splitDAO()} function in Figure \ref{fig:TheDAO}, which shows how the attack exploits the {\em reentrancy} vulnerability ($\V_{1}$). Specifically, the attack proceeds as follows: When a minority investor who opposes the proposal requests for a ``refund'', the {\tt DAO} contract creates a new {\tt DAO} contract account (Step {1-2}) and transfers the requesting investor's money to the new {\tt DAO} contract account (Step 3-4). The requesting investor may receive a reward for its past contribution or activity (Step 5-6). The vulnerability is that the number of tokens ``refunded'' to the requesting investor depends on the state variables {\myfont balances[msg.sender]} and {\myfont
totalSupply}, which are updated in the end of the {\sf splitDAO()} function after the external call {\sf msg.sender.call.value()}. This allowed the attacker to recursively call the {\sf splitDAO()} function (Step {7}) before the state variables are updated, causing a malicious investor to draw more money than it deserves \cite{DAOexploit}.

\subsubsection{Parity multisignature wallet attacks  (${\cal A}_2$ and ${\cal A}_3$)} 
In Ethereum, a multisignature wallet is a smart contract which requires multiple private keys to unlock a wallet in order to safeguard Ether or tokens. As shown in Figure \ref{fig:Parity}, a multisignature wallet supported by the Parity client consists of two contracts: (i) a library contract called {\tt WalletLibrary}, which implements all of the core functions of a wallet; and (ii) an actual {\tt Wallet} contract, which holds a reference (i.e., {\myfont \_walletLibrary}) that forwards all of the unmatched function calls to the library contract via {\sf delegatecall} (Line 7). The Parity multisignature wallet was compromised twice in 2017, which are briefly reviewed below.

\begin{figure}[!htbp]
\setlength{\abovecaptionskip}{0.cm}
\setlength{\belowcaptionskip}{-0.2cm}
\begin{lstlisting}[language=Solidity]
contract Wallet {
   address _walletLibrary = new WalletLibrary();
   address owner;
   ...
   function() payable {
      if (msg.data.length > 0)
         _walletLibrary.delegatecall(msg.data);
  }
}

contract WalletLibrary {
   ...
   function initWallet(address[] _owners, uint _required, uint _daylimit){                         
      initDaylimit(_daylimit);
      initMultiowned(_owners, _required);
   }
}
\end{lstlisting}
\caption{Simplified vulnerable Parity multisignature wallet.}
\label{fig:Parity}
\end{figure}

The first attack, denoted by $\A_2$, exploited the {\em delegatecall injection} vulnerability ($\V_{2}$) and the {\em erroneous visibility} vulnerability ($\V_{9}$) to drain Ethers approximately worthy of US\$31M. The attacker took over the ownership of contract {\tt Wallet} by sending a transaction to the contract with {\myfont msg.data} containing {\sf initWallet()} as the callee function (Line 13). Since contract {\tt Wallet} did not provide a function named {\sf initWallet()}, the contract's fallback function was triggered to delegate the wallet initialization task to the function {\sf initWallet()} in {\tt WalletLibrary}, which replaced the original multi-owner of contract {\tt Wallet} with the attacker's address specified in {\myfont msg.data}. This attack succeeds when the following four conditions are satisfied simultaneously \cite{parity-hack,parity-hack2}: (i) the function {\sf initWallet()} in the library was not specified as an {\em internal} one, meaning that it can be externally called via {\sf delegatecall}; (ii) the {\tt WalletLibrary} was actually a stateful contract, meaning that it can change the state of {\tt Wallet}; (iii) the function {\sf initWallet()} did not check whether the {\tt Wallet} contract had already been initialized (if so, no more initialization should be done); (iv) the {\tt Wallet}'s fallback function did not check the function being called, but unconditionally forwarded any unmatched calldata to {\tt WalletLibrary}, allowing unintended invocations.

The second attack, denoted by $\A_3$, exploited the {\em unprotected suicide} vulnerability ($\V_{10}$) and the {\em frozen Ether} vulnerability ($\V_{3}$), freezing approximately US\$280M in the affected wallets forever \cite{The280ME45:online}.
In response to the first attack, the Parity developers added a modifier, {\myfont only\_uninitialized}, to protect function {\sf initWallet()} such that a re-initialization of {\tt Wallet} via {\sf delegatecall} will throw an exception and be rejected by the modifier \cite{parity-attack2}. However, the shared {\tt WalletLibrary} itself was left uninitialized, which allowed an attacker to bypass the {\myfont only\_uninitialized} modifier and set himself as the owner of the {\tt WalletLibrary} \cite{destefanis2018smart}. Once taking over the library, the attacker invoked the {\sf suicide} method to kill the library, causing all of the {\tt Wallet} contracts relying on the library unusable.

\subsubsection{BECToken attack (${\cal A}_4$)} 
BECToken, an ERC-20 contract, was attacked in April 2018 by an exploitation of the {\em integer overflow} vulnerability ($\V_{6}$), causing a massive amount of tokens stolen and a temporary shutdown of token trading at exchange \cite{BatchOverflow}. The vulnerability was in the function {\sf batchTransfer()}
shown in Figure \ref{fig:BECToken}, and the function was meant for users to transfer tokens to multiple recipients via two arguments, one specifying the array of the recipients' addresses and the other specifying the respective number of tokens. The statement at Line 3 calculates the total number of tokens the sender should pay for a particular transaction, but may have the following integer overflow: By setting {\myfont \_value} to $2^{255}$ and {\myfont \_receivers} to two accounts controlled by the attacker, the attack overflows the 256-bit variable {\myfont amount} and makes it zero \cite{li2018detecting}. As a consequence, the attack bypasses the two checks at Lines 4-5 and causes sending the two receivers extremely large numbers of tokens.

\begin{figure}[!htbp]
\begin{lstlisting}[language=Solidity]
function batchTransfer(address[] _receivers, uint256 _value) public whenNotPaused returns (bool) {
   uint cnt = _receivers.length;
   uint256 amount = uint256(cnt) * _value;
   require(cnt > 0 && cnt <= 20);
   require(_value > 0 && balances[msg.sender] >= amount);
   ... // transfer to specified recipients
}
\end{lstlisting}
\caption{The vulnerable function in BECToken.}
\label{fig:BECToken}
\end{figure}

\subsubsection{GovernMental attacks (${\cal A}_5$,  ${\cal A}_6$, ${\cal A}_7$, and ${\cal A}_8$)} 
The contract {\tt GovernMental} was an array-based pyramid Ponzi scheme, where the last participant wins a jackpot if no one joins the scheme within 12 hours after the last participant \cite{bartoletti2017dissecting}. The contract has four vulnerabilities \cite{atzei2017survey}, which allowed the following four attack tactics and explains why ${\cal A}_5$,  ${\cal A}_6$, ${\cal A}_7$, and ${\cal A}_8$ belong to the same type of attacks. The first attack, denoted by $\A_5$, exploits the {\em DoS with unbounded operations}  vulnerability ($\V_{14}$) that when the array bookkeeping the number of participants becomes too large, the amount of gas required for operating on the array will go beyond the maximum gas that is permitted for assembling a block. This effectively halts the transaction and the winner cannot receive the 1,100 ETH jackpot.
The second attack, denoted by $\A_6$, exploits the {\em unchecked call return value} vulnerability ($\V_{15}$) that the contract does not check the returned value when sending profits to the winner and the {\em call-stack depth limit} vulnerability ($\V_{22}$). As a consequence, the owner of contract {\tt GovernMental} to fail the payment as follows: (i) Calling 1024 contracts before calling the target callee contract of the payee, which causes the callee contract to return {\sc False} to the caller contract, meaning that the callee contract does not receive the payment. (ii) The caller contract is supposed to check this returned value and then proceeds correspondingly but it does not. This causes the callee contract to lose money, which now belongs to the owner of the caller contract. The third attack, denoted by $\A_7$, exploits the {\em transaction-ordering dependence} vulnerability ($\V_{24}$) that a malicious miner can abandon some transactions related to {\tt GovernMental} or reorder transactions to make itself the last player (i.e., winner) in each round. 
The fourth attack, denoted by $\A_8$, exploits the {\em timestamp dependence} vulnerability ($\V_{25}$) that a malicious miner can manipulate {\myfont block.timestamp} so that its own block appears to be the last block to make itself win.

\subsubsection{HYIP attack (${\cal A}_9$)} 
The contract {\tt HYIP} was another Ponzi scheme, which pays existing investors from funds contributed by new investors at the end of each day. This mechanism is implemented by the function {\sf performPayouts()} highlighted in Figure \ref{fig:HYIP}, which contains the {\em DoS with unexpected revert} vulnerability ($\V_{5}$) (Line 7) \cite{bartoletti2017dissecting}.
The attack proceeds as follows: (i) The attacker, say Alice, writes an exploitation contract, named {\tt Mallory}, in which the attacker invests and throws an exception in the fallback function (Line 14). 
(ii) When function {\sf performPayouts()} is called to pay the investors, the fallback function is invoked and throws an exception, causing a reversion of the money transfer (Line 7) and thus DoS to contract {\tt HYIP}. (iii) The attacker can blackmail {\tt HYIP} to pay a ransom for halting its attack, by undoing the {\sf throw} operation (Line 15) via function {\sf stopAttack} (Line 18), which can only be done by the contract owner, Alice.

\begin{figure}[!htbp]
\begin{lstlisting}[language=Solidity]
contract HYIP {
   Investor[] private investors;
   ...
   function performPayouts() {
      for(uint idx = investors.length; idx-- > 0; ) {
         uint payout=(investors[idx].amount*33)/1000;
         if(!investors[idx].addr.send(payout)) throw;
      }
}}

contract Mallory {
   bool private attack = true;
   ...
   function() payable {
      if (attack) throw;
   }
   function stopAttack() {
      if(msg.sender == owner)  attack = false;
   }
}
\end{lstlisting}
\caption{Simplified HYIP contract and attack.}
\label{fig:HYIP}
\end{figure}

\subsubsection{Fomo3D attacks (${\cal A}_{10}$ and ${\cal A}_{11}$)} 
The contract {\tt Fomo3D} was an extremely popular Ponzi game in year 2018, where the last participant who buys a key before the timer runs out won the jackpot. The price of keys gradually grows with the number of buyers. When a key was sold, the countdown extends for 30 seconds. In addition to the jackpot winner, Fomo3D implemented an airdrop lottery to attract participants. For each purchase over 0.1 ETH, the participant (i.e., buyer) had a random chance to be picked up for a tiny profit from the prize pool. These two incentive mechanisms can be attacked as follows \cite{Fomo3D-1}.

The first attack, denoted by $\A_{10}$, is against the airdrop mechanism.
Specifically, the attack exploited the {\em generating randomness} vulnerability ($\V_{26}$). As shown in Figure \ref{fig:Fomo3D}, function {\sf airdrop()} generates a random seed by performing a deterministic computation on the current block state (i.e., {\myfont block.timestamp}, {\myfont block.difficulty}, etc.) and the address of {\myfont msg.sender} (Line 2-8). If the seed satisfies a certain condition (Line 10), the current key buyer wins an airdrop. However, since the block information is predictable, an attacker can simply pre-compute the addresses of new contracts and brute-forces for the winning seed (Line 2).

\begin{figure}[!htbp]
\setlength{\abovecaptionskip}{0.cm}
\setlength{\belowcaptionskip}{-0.3cm}
\begin{lstlisting}[language=Solidity]
function airdrop() private view returns(bool) {
   uint256 seed= uint256(keccak256(abi.encodePacked(
      (block.timestamp).add
      (block.difficulty).add
      ((uint256(keccak256(abi.encodePacked(block.coinbase)))) / (now)).add
      (block.gaslimit).add
      ((uint256(keccak256(abi.encodePacked(msg.sender)))) / (now)).add
      (block.number)
   )));
   if((seed-((seed/1000)*1000)) < airDropTracker_)
      return(true);
   else
      return(false);
   }
}
\end{lstlisting}
\caption{A snippet source code of Fomo3D.}
\label{fig:Fomo3D}
\end{figure}

The second attack, denoted by $\A_{11}$, is against the winning mechanism \cite{Fomo3D-1}. Specifically, the attack exploited the {\em DoS with block stuffing} vulnerability ($\V_{32}$) and causes the attacker to win the prize of approximately US\$3M \cite{Fomo3D-4}. The attack proceeds as follows: When the timer of the game reached about three minutes, the attacker bought a key and then sent multiple transactions to his own accounts with high enough {\em gasPrice}. Owing to the choice of miners, these transactions were first assembled into blocks. Since the maximum amount of gas consumption for a block is limited, any transactions related to {\tt Fomo3D} were not assembled into blocks. By congesting the network until the game was over, the attacker succeeded in becoming the last player.

\subsubsection{ERC-20 signature replay attack (${\cal A}_{12}$)} 
\begin{figure}[!htbp]
\setlength{\abovecaptionskip}{0.cm}
\setlength{\belowcaptionskip}{-0.3cm}
\includegraphics[width=.48\textwidth]{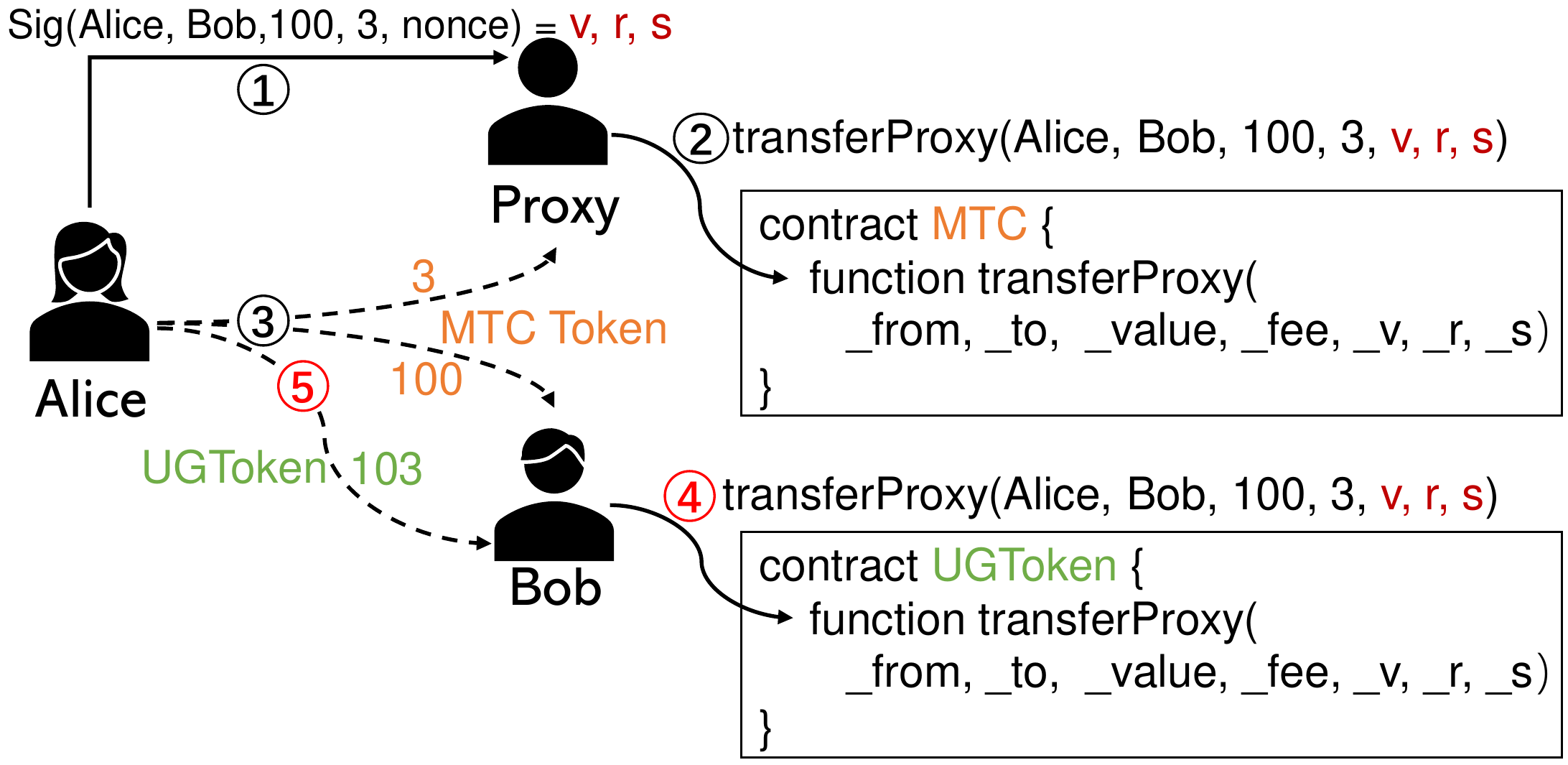}
\centering
\begin{lstlisting}[language=Solidity,basicstyle=\footnotesize ]
function transferProxy(address _from, address _to, uint256 _value, uint256 _fee, uint8 _v,bytes32 _r, bytes32 _s) returns (bool){
    if(balances[_from] < _fee + _value) throw;
    uint256 nonce = nonces[_from];
    bytes32 h = sha3(_from, _to, _value, _fee, nonce);
    if(_from != ecrecover(h, _v, _r, _s)) throw;
    if(balances[_to] + _value < balances[_to] || balances[msg.sender] + _fee < balances[msg.sender]) throw;
    balances[_to] += _value;
    Transfer(_from, _to, _value);
    balances[msg.sender] += _fee;
    Transfer(_from, msg.sender, _fee);
    balances[_from] -= _value + _fee;
    nonces[_from] = nonce + 1;
    return true;
}
\end{lstlisting}
\caption{Cross-contract replay attack via {\sf transferProxy()}.}
\label{fig:contract-replay-attack}
\end{figure}

This attack, which was first reported in \cite{Defcon-26}, exploits the 
{\em insufficient signature information} vulnerability ($\V_{13}$).
When a user transfers ERC-20 tokens, the user must have enough Ether to pay the transaction fee, which can be inconvenient when the user does not own any Ether. In order to alleviate the problem, the proxy-transfer method is introduced such that a user can authorize a proxy to carry out a transaction and pay the proxy some extra tokens as its service fee.
As shown in Figure \ref{fig:contract-replay-attack}, when Alice is to transfer 100 MTC tokens to Bob, she can send a signed message off-chain to a proxy (Step 1) such that the proxy launches a transaction to transfer 100 tokens to Bob and receives 3 token from Alice for the service (Step 2). A signature is verified using function {\sf transferProxy()}, which uses the Solidity function {\sf ecrecover()} (Line 5) to identify Alice's account address that issued the signature. However, Alice's off-chain message may not provide her token contract address, which should be bound to her signature. As a consequence, the signature can be accepted as valid with respect to any token contract address (e.g., MTC, UGToken, and GGoken), meaning that Bob can replay the signed message to other kinds of token contracts, such as UGToken (Step 4), to obtain extra money from Alice (Step 5) \cite{Replay-signatures}. 

\subsubsection{Rubixi attack (${\cal A}_{13}$)} 
The {\tt Rubixi} contract is a Ponzi scheme which contains the {\em erroneous constructor name} vulnerability ($\V_{17}$). The contract was originally named {\tt DynamicPyramid} and was later renamed by the developer to {\tt Rubixi}. However, the contract's constructor name was not updated accordingly, allowing anyone that calls the public function {\sf DynamicPyramid()} to become the owner of the contract and therefore steal the funds of the contract.

\subsection{Attacks at the data layer}
\subsubsection{ETH and ETC cross-chain replay attack (${\cal A}_{14}$)}
Ethereum had a hard fork after the DAO attack, splitting into ETH and ETC that share the same transaction history. This means that a transaction that was validated by the ETH network could also be accepted by the ETC network when the recipient immediately rebroadcast the transaction on the ETC network, and vice versa \cite{kiffer2017stick}. Since both ETC and ETH networks did not implement any defense against this attack that exploited the {\em indistinguishable chains}
vulnerability ($\V_{27}$), exchanges participating in both chains (e.g., Coinbase and Yunbi) lost a large amount of money \cite{Replay-coindesk}.

\subsubsection{Under-priced DDoS attacks (${\cal A}_{15}$ and ${\cal A}_{16}$)} 
These attacks, which were reported in \cite{spam-attack,Buterin,chen2017adaptive}, exploited both application-layer and data-layer vulnerabilities.

Specifically, the first attack, denoted by $\A_{15}$, exploited the {\em under-priced opcodes} vulnerability ($\V_{23}$) owing to the improper gas cost of EVM's {\sc extcodesize} opcode.
Prior to the EIP 150 hard fork, the {\sc extcodesize} opcode only charged 20 gas for reading a contract's bytecode from disk and then deriving its length. As a consequence, the attacker can repeatedly send transactions to invoke a deployed smart contract with many {\sc extcodesize} opcodes to cause a 2-3x slower block creation rate \cite{spam-attack}.

The second attack, denoted by $\A_{16}$, exploited the {\em under-priced opcodes} vulnerability ($\V_{23}$) owing to the improper gas cost of EVM's {\sc suicide} opcode
and the {\em empty account in the state trie} vulnerability ($\V_{28}$). The {\sc suicide} opcode (renamed to {\sc selfdestruct} after EIP 6) is meant to remove a deployed contract from the blockchain and send the remaining balance of Ether to the account designated by the caller. When the target account does not exist,  
a new account is created even though no Ether may be transferred; this consumes merely 90 gas \cite{Buterin}. Since an existing empty account is stored in the Ethereum state trie, the attacker created 19 million new empty accounts via the {\sc suicide} opcode at a low gas consumption, which wasted considerable disk space while increasing the synchronization and transaction processing time.

\subsection{Attacks at the consensus layer}
\subsubsection{Ethereum Classic (ETC) 51\% attack (${\cal A}_{17}$)} 
In January 2019, ETC suffered from a 51\% attack that exploited the {\em 51\% hashrate} vulnerability ($\V_{30}$), in which the attacker carried out double-spending transactions against several exchanges, causing an estimated loss of US\$1.1M \cite{ETC}. Since 2018, ETC's mining hashrate had dropped significantly due to the declining in its price, which lowered the amount of computing resources that are required for launching a 51\% attack. Moreover, cloud mining services (e.g., NiceHash) make it even easier to launch 51\% attacks. The attack was disrupted when exchanges increased the number of blocks that are required for transaction confirmation and limited the participation of malicious addresses in ETC trade.

\subsubsection{Selfish-mining attack (${\cal A}_{18}$)} 
The selfish-mining attack, which exploits the {\em rewards for uncle blocks} vulnerability ($\V_{33}$), means that miners may withhold their newly mined blocks and selectively publish some of their blocks to earn an unfair share of reward. A selfish miner continuously monitors the situation on a blockchain's public branches, estimates its advantages, and reveals its private blocks accordingly. When the public branches are shorter than the selfish-miner's private branch, the honest miners will switch to latter, rendering their previous mining effort  useless and making the selfish-miners receive a higher reward. 
Ritz et al. \cite{ritz2018impact} conducted a Monte Carlo simulation to emulate the block generation process in Ethereum and quantified the impact of uncle rewards on selfish-mining. Their simulation results showed that the uncle-block reward mechanism not only lowered the threshold of computational power at which selfish-mining becomes profitable, but also weakened the overall resilience against other attackers such as {\em double-spending}. Niu et al. \cite{niu2019selfish} developed a mathematical analysis on a selfish-mining strategy through a 2-dimensional Markov process model, showing that Ethereum is more vulnerable to selfish-mining than Bitcoin.

\subsubsection{Balance attack (${\cal A}_{19}$)} 
This attack exploits the {\em fixed consensus termination} vulnerability ($\V_{31}$) owing to the probabilistic PoW consensus in the presence of communication delays (or asynchronous networks) and the {\em unreliable BGP messages} vulnerability ($\V_{43}$).
This attack was first reported in \cite{natoli2017balance} via a theoretic analysis and testnet demonstration. The attack is to transiently partition the network into multiple subgroups of similar mining power so as to launch the double-spending attack at a subgroup of lower mining power. This allows the attacker to initiate transactions with merchants in one subgroup, while mining blocks in another subgroup to make its subtree outweighs the subtree mined in the victim group. After transactions in the victim subgroup get confirmed, the attacker reconnects the network. Since the mining power is roughly equally distributed among the subgroups, the subtree broadcast by the attacker has a good chance to be selected as the main chain, meaning that the attacker can double-spend in the victim subgroup. This attack was later deemed only theoretically possible because partitioning a public Ethereum network  (e.g., using BGP-hijacking) may not be feasible in practice \cite{ekparinya2018impact}. 

\subsection{Attacks at the network layer}

\subsubsection{Account hijacking attack (${\cal A}_{20}$)}
This attack exploits the {\em RPC API exposure} vulnerability ($\V_{38}$). In order to sign transactions, an EOA must first decrypt its private key that is stored on the local host and encrypted with a passphrase. This can be achieved by using the {\sf unlockAccount()} API of an Ethereum client, which uses the passphrase to obtain the private key and loads it into the memory for 300 seconds (by default). The private key in memory can be accessed by any Ethereum API without authentication. This can be exploited as follows.
Ethereum clients (e.g., Geth, Parity) typically use the default ports 8545 (HTTP) and 8546 (WebSocket) for the JSON-RPC interface. However, these client software neither configure those ports as local-only by default, nor adopt precautions (e.g., disabling remote calls). This allows an attacker to scan open Ethereum nodes and invoke {\sf eth\_sendTransaction()} API to transfer victims' money to the attacker's account. Once a victim types its passphrase to unlock its account, the  {\sf eth\_sendTransaction()} API will be successfully executed. By the time the attack was observed in March 2018 \cite{Security_RPC}, attackers had already stolen around US\$20M from exposed Ethereum clients.

\subsubsection{Eclipse attacks ({${\cal A}_{21}$ and ${\cal A}_{22}$})}
This attack allows an attacker, who can hijack the connections of some victim nodes in the P2P network, to create a virtual partition to isolate those victim nodes from the rest of the network. 
Victim nodes' connections can be hijacked at least by connection monopolization and poisoning victims' routing tables \cite{marcus2018low}.
The connection monopolization attack, denoted by $\A_{21}$, exploits the {\em unlimited nodes creation} vulnerability ($\V_{34}$) and the {\em uncapped incoming connections} vulnerability ($\V_{35}$). When a client reboots, it has no incoming or outgoing connections. An attacker can create plenty of node IDs in advance and initiate enough incoming connections to the victim node immediately after its reboot. A node is eclipsed when its connection slots (25 by default) are occupied by incoming connections from the attacker.
Despite the response to the aforementioned connection monopolization attack in imposing an upper limit on the number of incoming TCP connections, the following attack, denoted by $\A_{22}$, still can succeed. Specifically, an attacker still can exploit the {\em public peer selection} vulnerability ($\V_{36}$) to
poison the victim nodes' routing tables when these tables are reboot and reset (e.g., the attacker could craft fake nodes and insert them into those routing tables to make the victim nodes' outgoing TCP connections point to the fake nodes controlled by the attacker). The attacker can further occupy the victim nodes' remaining connection slots by initiating connections to the victim nodes.

\subsection{Attacks against the environment}

\subsubsection{EtherDelta attack (${\cal A}_{23}$)} 

EtherDelta is a popular exchange for users to trade ERC-20 tokens. It suffered from a code injection or EtherDelta attack that exploited the {\em Cross-Site Scripting} vulnerability ($\V_{40}$) in September 2017, causing a loss in cryptocurrency worth of thousands of dollars \cite{EtherDelta}. 
The attacker constructed a new token contract, which contains a piece of malicious JavaScript code in the token's name. Recall that EtherDelta extracts a newly created token's name from the token contract's code and displays the name on the exchange's website. Recall also that when performing a token-trade transaction, a user needs to load its private key and account address to the web browser for signing the transaction. As a consequence, when the name of a newly created token  was displayed on the user's browser, the malicious JavaScript code was executed to steal the user's private key from the browser, causing the loss of money protected by the private key.

\subsubsection{CoinDash attack (${\cal A}_{24}$)} 
CoinDash, a portfolio management platform, was compromised due to an exploitation of the {\em broken access control} vulnerability
($\V_{42}$) during the course of its ICO in Jul 2017, causing the loss of US\$7M worth of Ether within a few minutes. The attack proceeds by getting into the infected web application that hosts CoinDash's webpage, and replacing the ICO contract address with one that is controlled by the attacker \cite{CoinDash}. 

\subsubsection{Enigma attack (${\cal A}_{25}$)} 
Enigma, a decentralized investment platform, was attacked prior to its ICO in August 2017 \cite{Enigma}, owing to the exploitation of the {\em weak password} vulnerability ($\V_{39}$). Through social engineering means, the attacker successfully stole an Enigma founder's password, which was disclosed in an unrelated attack and reused in Enigma. As a consequence, the attacker took control of the company’s Slack channel, email lists, and Google account hosting the ICO's presale. The attacker replaced the official ICO contract address with the attacker's own address and sent messages to solicit buyers in fake presales. 

\subsubsection{MyEtherWallet attack (${\cal A}_{26}$)} 
MyEtherWallet, a popular web-based wallet, was attack in April 2018 owing to the exploitation of the {\em unreliable BGP messages} vulnerability ($\V_{43}$), the {\em sensitive DNS} vulnerability ($\V_{44}$), and the {\em unvalidated URL redirection} vulnerability ($\V_{41}$).
The attack exploits a joint BGP and DNS hijacking to mislead users to a fake version of the website and compromised the victims' private keys \cite{MyEtherWallet}. The hacker then emptied the victims' wallets and stole approximately US\$17M. 
The attacker first exploited a routing weakness in BGP and hijacked the traffic to Amazon's Route 53 servers, which provided DNS service to MyEtherWallet. When users visit their MyEtherWallet, their requests were rerouted to the fake DNS servers controlled by the attacker, which returned IP addresses to  direct users to a phishing website. The unsuspecting users, who neglected the TLS/SSL certificate warning sign, proceeded to login to the phishing site, causing their passphrases and private keys exposed.

\subsection{Further analysis of attack consequences} 

\begin{figure*}[!htbp]
\centering
\includegraphics[width=.98\textwidth]{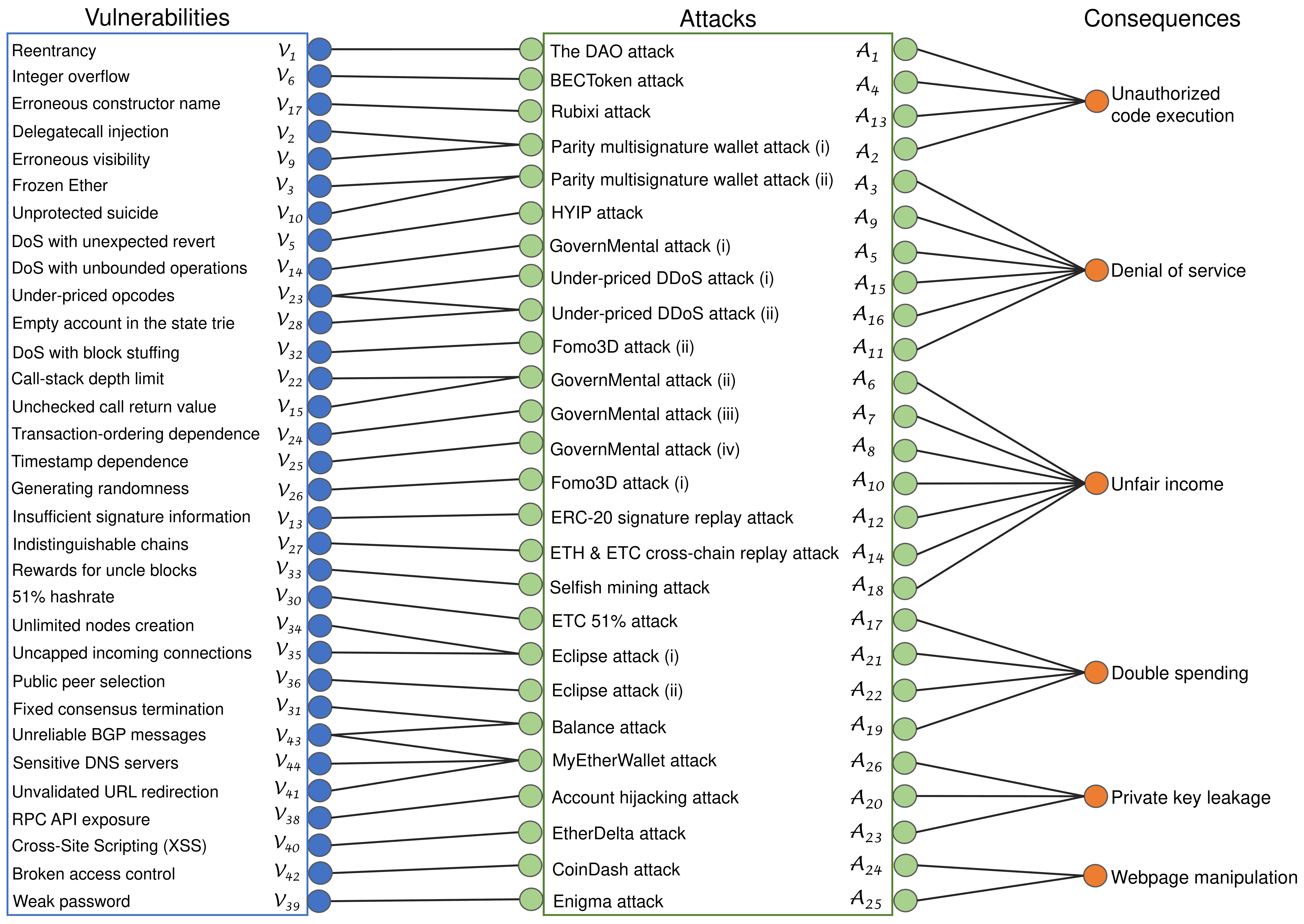}
\caption{The relation between vulnerabilities, attacks and attack consequences.
\label{fig:vuls-attack-consequence}}
\end{figure*}

Now we present a taxonomy of the attack consequences mentioned above: {\em unauthorized code execution}, {\em DoS}, {\em unfair income}, {\em double-spending}, {\em private key leakage}, and {\em webpage manipulation}, which are elaborated below. 
Figure \ref{fig:vuls-attack-consequence} highlights the relationship between these attack consequences, attacks, and vulnerabilities, where
the vulnerabilities and attacks are ordered in such a way that allows us to clearly present the relationships between them (i.e., neither the vulnerabilities nor the attacks are ordered as they are discussed in the text). For ease of reference, we use ``${\cal A}_i({\cal V}_j,\cdots)$'' to denote that attack ${\cal A}_i$ exploits vulnerabilities ${\cal V}_j$ and possibly others.

\subsubsection{Unauthorized code execution}
This consequence occurs in one of the following four scenarios. (i) The attacker uses the DAO attack (${\cal A}_1$) to exploit the {\em reentrancy} vulnerability (${\cal V}_1$), namely ${\cal A}_1({\cal V}_1)$. 
(ii) The attacker uses the first Parity multisignature wallet attack (${\cal A}_2$) to exploit the {\em delegatecall injection} vulnerability (${\cal V}_2$) and the {\em erroneous visibility} vulnerability (${\cal V}_9$), namely ${\cal A}_2({\cal V}_2, {\cal V}_9)$.
(iii) The attacker uses the BECToken attack (${\cal A}_4$) to exploit the {\em Integer overflow} (${\cal V}_6$), namely ${\cal A}_4({\cal V}_6)$.
(iv) The attacker wages the Rubixi attack (${\cal A}_{13}$) to exploit the {\em erroneous constructor name} (${\cal V}_{17}$), namely ${\cal A}_{13}({\cal V}_{17})$. The biggest financial loss caused by a single unauthorized code execution is US\$60M, owing to the DAO attack (${\cal A}_{1}$) \cite{TheDAO}.

\subsubsection{DoS}
This consequence can be further divided into two sub-consequences: {\em DoS against smart contracts} and {\em DoS against Ethereum network}. 
The former sub-consequence occurs in one of the following four scenarios:
(i) The attacker wages the second Parity multisignature wallet attack ($\A_{3}$) to exploit the {\em frozen Ether} vulnerability (${\cal V}_{3}$) and the
{\em unprotected suicide} vulnerability (${\cal V}_{10}$), namely $\A_{3}(\V_3,\V_{10})$.
(ii) The attacker uses the HYIP attack ($\A_{9}$) to exploit the 
{\em DoS with unexpected revert} vulnerability (${\cal V}_{5}$), namely $\A_{9}(\V_5)$.
(iii) The attacker uses the GovernMental attack ($\A_{5}$) to exploit the {\em DoS with unbounded operations} vulnerability (${\cal V}_{14}$), namely $\A_{5}(\V_{14})$.
(iv) The attacker uses the Fomo3D attack ($\A_{11}$) to exploit the {\em DoS with block stuffing} vulnerability (${\cal V}_{32}$), namely $\A_{11}(\V_{32})$. 
The latter sub-consequence occurs in one of the following scenarios:
(i) The attacker uses the under-priced DDoS attack ($\A_{15}$) to exploit the {\em under-priced opcodes} vulnerability (${\cal V}_{23}$), namely $\A_{15}(\V_{23})$.
(ii) The attacker uses the under-priced DDoS attack ($\A_{16}$) to exploit the {\em under-priced opcodes} vulnerability (${\cal V}_{23}$) and the {\em empty account in the state trie} vulnerability (${\cal V}_{28}$), namely $\A_{16}({\V_{23}, \V_{28}})$.
The biggest financial loss caused by a single DoS attack is  US\$280M, owing to the second attack against the Parity wallet ($\A_{3}$) \cite{The280ME45:online}.

\subsubsection{Unfair income}
This consequence occurs in one of the following scenarios: (i) A malicious contract owner uses the GovernMental attack ($\A_{6}$) to exploit the {\em unchecked call return value} vulnerability (${\cal V}_{15}$) and the {\em call-stack depth limit} (${\cal V}_{22}$) vulnerability to hold the money that should be transferred to a callee contract or EOA, namely $\A_{6}({\V_{15}, \V_{22}})$. (ii) A malicious miner wages the GovernMental attack ($\A_{7}$) to exploit the {\em transaction-ordering dependence} vulnerability (${\cal V}_{24}$), namely $\A_{7}(\V_{24})$.
(iii) A malicious miner wages the GovernMental attack ($\A_{8}$) to exploit the
{\em timestamp dependence} vulnerability (${\cal V}_{25}$), namely $\A_{8}(\V_{25})$.
(iv) The attacker wages the Fomo3D attack ($\A_{10}$) to exploit the {\em generating randomness} vulnerability (${\cal V}_{26}$) to make itself the winner, namely $\A_{10}(\V_{26})$. (v) An attacker uses the ERC-20 signature replay attack ($\A_{12}$) to exploit the {\em insufficient signature information} vulnerability (${\cal V}_{13}$), namely $\A_{12}(\V_{13})$.
(vi) The attacker uses the ETH and ETC cross-chain replay attack ($\A_{14}$) to exploit 
the {\em indistinguishable chains} vulnerability (${\cal V}_{27}$) to repeatedly withdraw money, namely $\A_{14}(\V_{27})$. 
(vii) A malicious miner wages the selfish mining attack ($\A_{18}$) to exploit the {\em rewards for uncle blocks} vulnerability (${\cal V}_{33}$) to get a higher mining reward, namely $\A_{18}(\V_{33})$.
The biggest financial loss caused by a single unfair income attack is US\$0.5M, owing to an ETH and ETC cross-chain replay attack ($\A_{14}$) \cite{Replay-coindesk}.

\subsubsection{Double-spending} 
This consequence occurs in one of the following four scenarios: (i) The attacker uses the 51\% attack ($\A_{17}$) to exploit the {\em 51\% hashrate} vulnerability (${\cal V}_{30}$), namely $\A_{17}(\V_{30})$.
(ii) The attacker uses the eclipse attack ($\A_{21}$) to exploit the {\em unlimited nodes creation} vulnerability (${\cal V}_{34}$) and the {\em uncapped incoming connections} vulnerability (${\cal V}_{35}$), namely $\A_{21}({\V_{34}, \V_{35}})$.
(iii) The attacker uses the eclipse attack ($\A_{22}$) to exploit the {\em public peer selection} vulnerability (${\cal V}_{36}$), namely $\A_{22}(\V_{36})$.
(iv) The attacker uses the balance attack ($\A_{19}$) to exploit the {\em fixed consensus termination} vulnerability (${\cal V}_{31}$) and the {\em unreliable BGP messages} vulnerability (${\cal V}_{43}$), namely $\A_{19}({\V_{31}, \V_{43}})$. The biggest financial loss caused by a single double-spending attack is about US\$1.1M, owing to the ETC 51\% attack ($\A_{17}$) \cite{ETC}.

\subsubsection{Private key leakage}
This consequence occurs in one of the following three scenarios.
(i) The attacker uses the MyEtherWallet attack (${\cal A}_{26}$) to exploit the {\em unvalidated URL redirection} vulnerability (${\cal V}_{41}$), the {\em unreliable BGP messages} vulnerability (${\cal V}_{43}$), and the {\em sensitive DNS} vulnerability (${\cal V}_{44}$) to direct the victim to a malicious website, namely ${\cal A}_{26}({\cal V}_{41}, {\cal V}_{43}, {\cal V}_{44})$. 
(ii) A user's private key is stolen by an attacker that uses the account hijacking attack (${\cal A}_{20}$) to exploit the {\em RPC API exposure} vulnerability (${\cal V}_{38}$) to hijack the victim's account from a remote Ethereum client, namely ${\cal A}_{20}({\cal V}_{38})$.
(iii) The attacker uses the EtherDelta attack (${\cal A}_{23}$) to exploit the {\em Cross-Site Scripting} vulnerability (${\cal V}_{40}$) to hijack the victim's browser session, namely ${\cal A}_{23}({\cal V}_{40})$. The biggest financial loss caused by a private key leakage is US\$20M, owing to an account hijacking attack (${\cal A}_{20}$) \cite{Security_RPC}.

\subsubsection{Webpage manipulation}
This consequence occurs in one of the following two scenarios.
(i) A DApp's website interface is compromised by an attacker that wages the CoinDash attack (${\cal A}_{24}$) to
exploit the {\em broken access control} vulnerability (${\cal V}_{42}$), namely ${\cal A}_{24}({\cal V}_{42})$.
(ii) The attacker uses the Enigma attack (${\cal A}_{25}$) to exploit the {\em weak password} vulnerability (${\cal V}_{39}$), namely ${\cal A}_{25}({\cal V}_{39})$. These vulnerabilities reside in, or are inherited from, the Ethereum environment. The biggest financial loss caused by a webpage manipulation is US\$7M, owing to an attack against CoinDash (${\cal A}_{24}$) \cite{CoinDash}.

\subsubsection{Summary and insights}

\begin{table}[!htbp]
  \centering
  \scriptsize
  \setlength\tabcolsep{2pt}
  \caption{Overview of attacks and financial losses.}
    \begin{tabular}{|p{3.7em}|p{5em}|l|l|p{4em}|}
    \hline
    Attack targets & Vulnerability location &  \multirow{1.5}[3]*{Real-world attacks} & \multirow{1.5}[3]*{Attack consequences} & Financial losses \bigstrut\\
    \hline
    \multirow{7}[14]{*}{DApp} & \multirow{3}[6]{*}{Application} & The DAO ($\A_1$) & Unauthorized code exec. & US\$60M \bigstrut\\
\cline{3-5}      &   & Parity (i) ($\A_2$) & Unauthorized code exec. & US\$31M \bigstrut\\
\cline{3-5}      &   & Parity (ii) ($\A_3$) & DoS & US\$280M \bigstrut\\
\cline{2-5}      & \multirow{3}[6]{*}{Environment} & CoinDash ($\A_{24}$) & Webpage manipulation & US\$7M \bigstrut\\
\cline{3-5}      &   & Enigma ($\A_{25}$) & Webpage manipulation & US\$0.5M \bigstrut\\
\cline{3-5}      &   & MyEtherWallet ($\A_{26}$) & Private key leakage & US\$17M \bigstrut\\
\cline{2-5}      & \multirow{2}[4]{*}{Consensus} &Fomo3D (ii) ($\A_{11}$)& DoS & US\$3M \bigstrut\\
\cline{1-1}\cline{3-5}    \multirow{3}[6]{*}{Ethereum } &   & ETC 51\%  ($\A_{17}$) & Double spending & US\$1.1M \bigstrut\\
\cline{2-5}      & \multirow{1}[2]{*}{Data} & ETH \& ETC replay ($\A_{14}$) & Unfair income & US\$0.5M \bigstrut\\
\cline{2-5}      & Network  & Account hijacking ($\A_{20}$) & Private key leakage & US\$20M \bigstrut\\
    \hline
    \end{tabular}%
  \label{tab:attack-consequence-summary}
\end{table}%

Table \ref{tab:attack-consequence-summary}
summarizes the aforementioned attacks, the layers at which the vulnerabilities they exploit reside, attack consequences, and financial losses incurred by those attacks. This allows us to draw the following insights:

\begin{insight}
\label{insight-dos-attack}
The biggest financial loss in the Ethereum system is \$280M, which is caused by a DoS attack against the Parity wallet that disabled a library that is used by many contracts. 
\end{insight}

Insight \ref{insight-dos-attack} is interesting at least from two points of view. First, DoS often causes {\em indirect} economic losses incurred by the unavailability of services. The DoS attack against Ethereum manifests {\em direct} economic losses. Second, the DoS attack is  caused by code-reuse, which is a widely adopted practice in the traditional paradigm of programming. However, in the paradigm of DApps, code-reuse may impose a higher risk than ever, highlighting the importance of security auditing for popular contracts as well as the libraries they use.

\begin{insight}
\label{insight-decentralization}
DApps running on top of the Ethereum blockchain are not fully {\em decentralized} when they use centralized  web interfaces.
\end{insight}

Insight \ref{insight-decentralization} is interesting because it points out one scenario that manifests that blockchain-based systems are {\em not} necessarily decentralized, despite the popular view. Indeed, vulnerabilities in those web interfaces can cause large financial losses. This highlights the importance of decentralizing the web interfaces to the blockchain-based systems.

\begin{insight}
\label{insight-attack-application-layer-importance}
Application-layer attacks have caused the largest financial losses among the attacks against Ethereum.
\end{insight}

Insight \ref{insight-attack-application-layer-importance} is interesting because a vulnerability at a lower layer would often cause a larger damage than a vulnerability at a higher layer. This rule of thumb does not necessarily apply to Ethereum or blockchain-based systems in general. This can be (at least partly) attributed to the fact that DApps operate directly on the digital assets, and highlights the importance of assuring application-layer security.

\section{Defenses}
\label{sec:defenses}

In this section, we describe the 47 defenses that have been proposed for securing the Ethereum ecosystem, which are denoted by ${\cal D}_1,\ldots,{\cal D}_{47}$, respectively. Unlike vulnerabilities and attacks which naturally correspond to some layer of the Ethereum architecture, defenses are by no means geared towards the layers. Therefore, we propose categorizing them into two classes: proactive defenses and reactive defenses.

\subsection{Proactive defenses}
We categorize existing proactive defense mechanisms into the following five sub-classes based on the respective focuses of these mechanisms: {\em contract programming language}, {\em contract development}, {\em contract analysis}, {\em contract and Ethereum enhancement}, and {\em consensus protocols}.

\subsubsection{Language-based security}
Programming language approaches to securing smart contracts can be divided into two categories: high-level languages for developing more secure smart contracts and intermediate-level languages for facilitating contract formal analysis. 

\paragraph{High-level languages}
Vyper \cite{Vyper:online} (${\cal D}_1$) removes a number of functionalities provided by Solidity (e.g., recursive calling, infinite loops, modifiers) and adds several new features (e.g., bounds and overflow checking) to eliminate vulnerabilities, such as the {\em DoS with unbounded operations} vulnerability ($\V_{14}$), the {\em erroneous visibility} vulnerability ($\V_{9}$), and the {\em Integer overflow and underflow} vulnerability ($\V_{6}$). Bamboo \cite{GitHubpi3:online} (${\cal D}_2$) uses polymorphism to mitigate the 
{\em transaction-ordering dependence} vulnerability ($\V_{24}$), while eliminating the {\em reentrancy} vulnerability ($\V_{1}$) and the {\em DoS with unbounded operations} vulnerability ($\V_{14}$). Obsidian \cite{coblenz2017obsidian} (${\cal D}_3$) models smart contracts as finite state machines and tracks contracts' states to eliminate the {\em reentrancy} vulnerability ($\V_{1}$). Flint \cite{schrans2018writing} (${\cal D}_4$) uses an {\myfont Asset} type to assure the atomicity of operations and introduces restrictions on callers' capabilities to protect contract functions from unauthorized access, while aiming to eliminate the {\em reentrancy} vulnerability ($\V_{1}$), the {\em erroneous visibility} vulnerability ($\V_{9}$), the {\em Integer overflow and underflow} vulnerability ($\V_{6}$), and the {\em timestamp dependency} vulnerability ($\V_{25}$).

\paragraph{Intermediate-level languages}
Simplicity \cite{o2017simplicity} (${\cal D}_5$) is an intermediate representation between high-level languages and low-level virtual machine. It provides formal semantics using the proof-assistant Coq \cite{WelcomeT3:online}, thus allowing formally analysis of contracts properties (e.g., safety and liveness). Scilla \cite{sergey2018scilla} (${\cal D}_6$) distinguishes  in-contract computation from inter-contract communication to disentangle contract-specific effects from each other.

\subsubsection{Contract development}
Since smart contracts are a new programming paradigm, these defenses can help developers in avoiding or mitigating common vulnerabilities.

\begin{figure*}[!htbp]
\includegraphics[width=.98\textwidth]{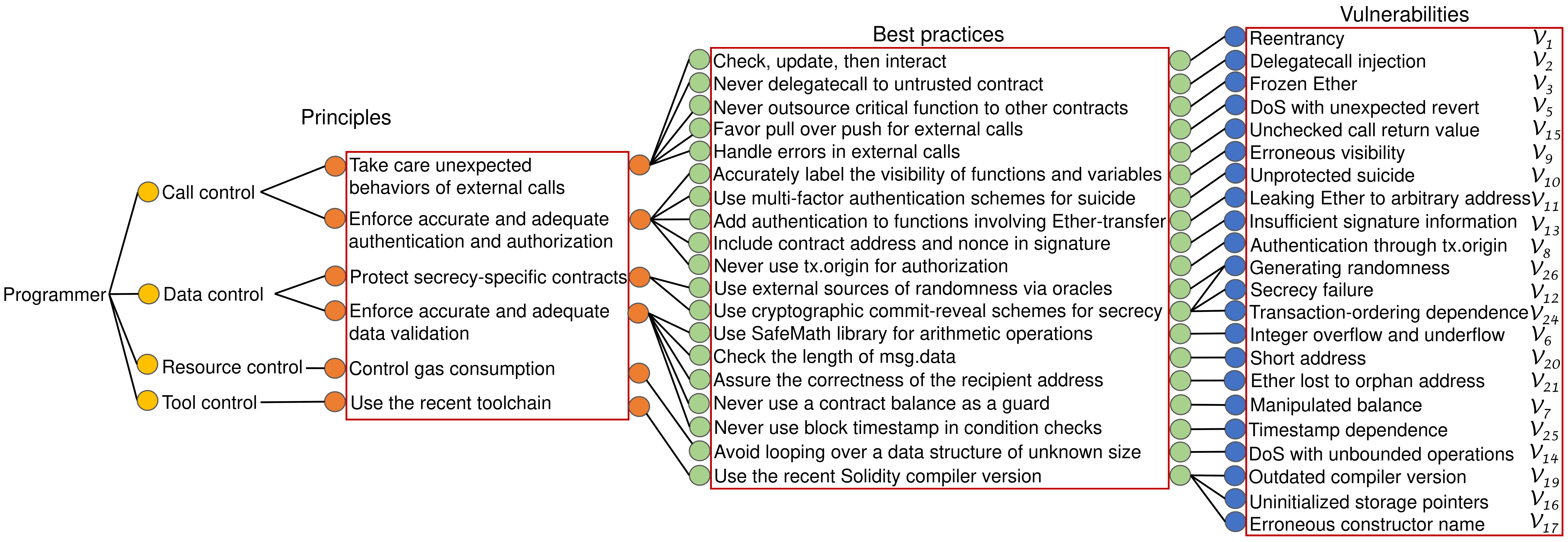}
\centering
\caption{Best practices and principles for guiding contract development.
\label{fig:best-practices}}
\end{figure*}

\paragraph{Principles and best practices}
A number of Solidity best practices were recommended in \cite{Solidity80:online}, such as  {\em check, update, then interact} (i.e., checking conditions first, then updating state variables, and finally interacting with other contracts); {\em favor pull over push for external calls} (i.e., letting a recipient withdraw or ``pull'' the money set by the sender, rather than letting the sender directly transfer or ``push'' the money to the recipient). 
In order to make it easier for practitioners 
to adopt those best practices, we systematize 19 best practices (${\cal D}_7$) according to the vulnerabilities towards which they are geared, while noting that 15 of these 19 best practices were scattered in \cite{Solidity80:online,SWC,Adrian_list}. Our systematization leads to 4 general principles or 6 specific sub-principles in total.

Figure \ref{fig:best-practices} highlights these (sub-)principles.  Intuitively, a programmer should be conscious of four kinds of controls: {\em call control}, {\em data control}, {\em resource control}, and {\em tool control}. 
(i) The {\em call control} principle says that a programmer should secure the interactions between smarts contracts and the interactions between EOAs and smart contracts. This principle is further divided into two sub-principles: one coping with the callee's unexpected behaviors and the other coping with the caller's access control.
(ii) The {\em data control} principle says that a programmer should secure the data flow of a contract. This principle is further divided into two sub-principles: one dealing with the protection of sensitive data and the other dealing with the prevention of malformed data from entering a smart contract. 
(iii) The {\em resource control} principle says that a programmer should mitigate the exhaustion of available gas in Ethereum. 
(iv) The {\em tool control} principle says that a programmer should use updated tools (e.g., compiler, debugger) to eliminate known vulnerabilities.

\paragraph{Software engineering mechanisms}
In order to defend against attacks that may exploit unknown vulnerabilities, several blockchain-oriented software engineering mechanisms were introduced \cite{software_engineering} (${\cal D}_8$), such as: the {\it rate limit} mechanism, which restricts the number of consecutive actions incurred by an EOA or restricts the amount of Ether sent by a contract within a period of time; the {\it balance limit} mechanism, which regulates the maximum amount of funds that can be held by a contract; the {\it speed bump} mechanism, which postpones some potentially-damaging operations.

\subsubsection{Smart contract analysis}
These defenses aim to enhance security of smart contracts via the following approaches: symbolic execution, abstract interpretation,
formal verification, fuzzing and model-based vulnerability detection.

\paragraph{Symbolic execution}
It works on a program' control-flow graph (CFG) and traverses all of the feasible execution paths on the graph to identify vulnerabilities \cite{king1976symbolic}. 
This approach achieves neither soundness (i.e., zero false-negatives) nor completeness (i.e., zero false-positives), owing to the omission of some execution paths and the exploration of unreachable paths in real executions. Oyente \cite{luu2016making} (${\cal D}_9$) can detect four types of vulnerabilities --- the {\em reentrancy} vulnerability ($\V_{1}$), the {\em mishandled exceptions} vulnerability ($\V_{15}$), the {\em transaction-ordering dependence} vulnerability ($\V_{24}$), and the {\em timestamp dependence} vulnerability ($\V_{25}$) --- but incurs a high false-positive rate \cite{kalra2018zeus}. Maian \cite{nikolic2018finding} (${\cal D}_{10}$) extends Oyente by considering multiple invocations of a contract, rather than a single invocation. It can detect three types of vulnerabilities: the {\em frozen Ether} vulnerability ($\V_{3}$), the {\em leaking Ether to arbitrary address} vulnerability ($\V_{11}$), and the {\em unprotected suicide} vulnerability ($\V_{10}$). 
Mythril \cite{Mythril} (${\cal D}_{11}$) uses ``concolic analysis'', which integrates symbolic and concrete execution of a smart contract, to detect 8 types of vulnerabilities, such as the {\em manipulated balance} vulnerability ($\V_{7}$), the {\em authentication through tx.origin} vulnerability ($\V_{8}$), and the {\em generating randomness} vulnerability ($\V_{26}$).
The teEther tool \cite{krupp2018teether} (${\cal D}_{12}$) can detect vulnerabilities like the {\em erroneous visibility} vulnerability ($\V_{9}$) and the {\em erroneous constructor name} vulnerability ($\V_{17}$).
The sCompile tool \cite{chang2018scompile} (${\cal D}_{13}$) reduces the number of suspicious execution paths that are not related to  money-transfer. It can detect 4 types of vulnerabilities: the {\em reentrancy} vulnerability ($\V_{1}$), the {\em frozen Ether} vulnerability ($\V_{3}$), the {\em unprotected suicide} vulnerability ($\V_{10}$), and the {\em Ether lost to orphan address} vulnerability ($\V_{21}$).
ECF \cite{grossman2017online} (${\cal D}_{14}$) focuses on detecting 
callback-related vulnerabilities, such as the {\em reentrancy} vulnerability ($\V_{1}$).

\begin{figure}[!htbp]
\setlength{\abovecaptionskip}{0.2cm}
\setlength{\belowcaptionskip}{-0.3cm}
\includegraphics[width=.45\textwidth]{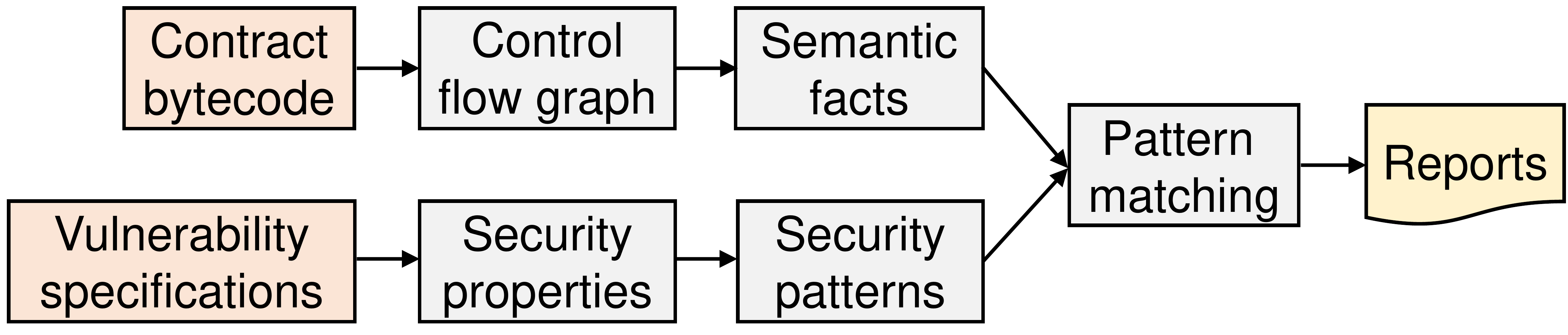}
\centering
\caption{Workflow of vulnerability detector Securify.}
\label{fig:interpretation}
\end{figure}

\paragraph{Abstract interpretation}
Abstract interpretation aims to over-approximate the semantics of a program so as to achieve soundness in program analysis \cite{cousot1977abstract}. Securify \cite{tsankov2018securify} (${\cal D}_{15}$) defines a set of {\it compliance} and {\it violation} patterns to characterize how contract comply and violate security properties extracted from some known vulnerabilities, such as the {\em reentrancy} vulnerability ($\V_{1}$), the {\em delegatecall injection} vulnerability ($\V_{2}$), and the {\em frozen Ether} vulnerability ($\V_{3}$). Figure \ref{fig:interpretation} highlights how these patterns can be used to detect vulnerabilities. Zeus \cite{kalra2018zeus} (${\cal D}_{16}$) defines a set of correctness and fairness policies and then embeds them as {\myfont assert} statements into the source code of contracts. It can detect six types of vulnerabilities, such as the {\em reentrancy} vulnerability ($\V_{1}$) and the {\em unchecked call return value} vulnerability ($\V_{15}$).  
FSolidM \cite{mavridou2017designing} (${\cal D}_{17}$) 
abstracts smart contracts as finite state machines and can detect the {\em reentrancy} vulnerability ($\V_{1}$) and the {\em transaction-ordering dependence} vulnerability ($\V_{24}$). MadMax \cite{grech2018madmax} (${\cal D}_{18}$) disassembles EVM bytecode into an intermediate representation and then leverages both data-flow analysis and context-sensitive flow analysis to detect gas-related vulnerabilities, such as the {\em DoS with unbounded operations} vulnerability ($\V_{14}$). Vandal \cite{brent2018vandal} (${\cal D}_{19}$) translates EVM bytecode into logic relations and use them to detect a number of vulnerabilities, such as the
{\em reentrancy} vulnerability ($\V_{1}$), the {\em authentication through tx.origin} vulnerability ($\V_{8}$), the {\em unprotected suicide} vulnerability ($\V_{10}$), and the {\em unchecked call return value} vulnerability ($\V_{15}$).

In order to facilitate the aforementioned high-level contract analysis, several reverse engineering tools have been developed to convert EVM bytecode to source code or intermediate presentation. Porosity \cite{suiche2017porosity} (${\cal D}_{20}$) is a decompiler for producing Solidity source code from EVM bytecode. Erays \cite{zhou2018erays} (${\cal D}_{21}$) lifts EVM bytecode to a high-level pseudocode by recovering the control-flow structure and transforming EVM from a stack-based machine to a register-based machine. EthIR \cite{albert2018ethir} (${\cal D}_{22}$) decompiles EVM bytecode to a high-level rule-based representation, which can then be fed into an automated static analyzer to infer high-level properties of the EVM bytecode.

\begin{figure}[!htbp]
\setlength{\abovecaptionskip}{0.2cm}
\setlength{\belowcaptionskip}{-0.3cm}
\includegraphics[width=.42\textwidth, height= .18\textwidth]{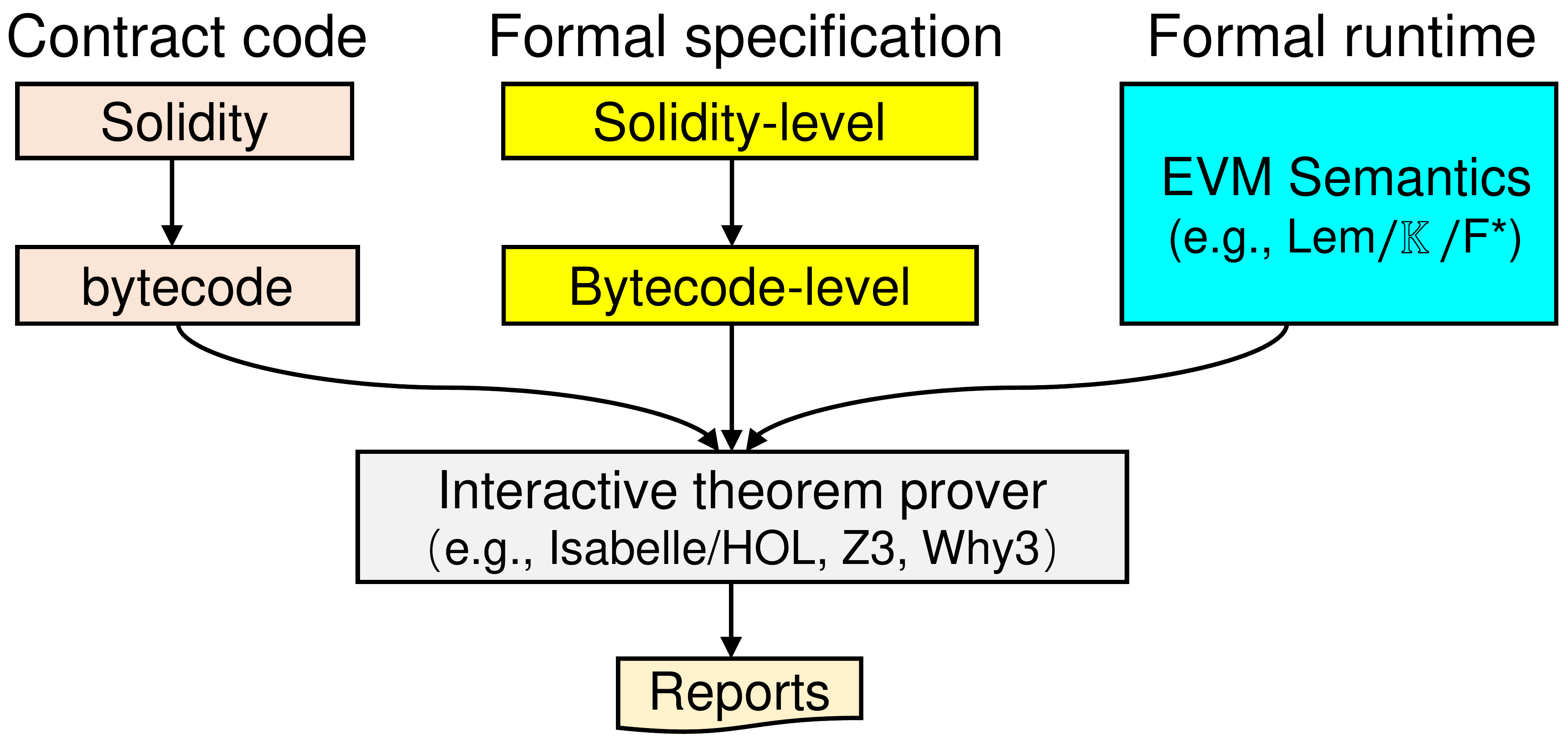}
\centering
\caption{Theorem proving for verifying smart contracts.
\label{fig:formal-verification}}
\end{figure}

\paragraph{Formal verification}
Formal verification proves the correctness of contract implementation with respect to a specification. 
This approach assures the completeness {(i.e., no false positives).}
Figure \ref{fig:formal-verification} illustrates
how to use theorem-proving to verify smart contracts at the EVM bytecode-level.
Hirai \cite{hirai2017defining} (${\cal D}_{23}$) took the first step towards formalizing the EVM semantics, which can be accommodated by interactive theorem provers like Isabelle/HOL \cite{nipkow2002isabelle} to prove invariants and safety properties of smart contracts. Amani et al. \cite{amani2018towards} (${\cal D}_{24}$) extend the work in \cite{hirai2017defining} by splitting a contract into some basic blocks and then using a Hoare-style program logic to reason about semantic properties of  contracts from properties of its parts.
Hildenbrandt et al. \cite{hildenbrandt2018kevm} (${\cal D}_{25}$) present a complete semantics of the EVM, referred to as KEVM, using the $\mathbb{K}$ framework \cite{rou2010overview} to achieve language-independent program verification. Park et al. \cite{park2018formal} (${\cal D}_{26}$) optimize the KEVM verifier by introducing EVM-specific abstractions and lemmas to avoid non-tractable reasoning in the underlying theorem prover. Grishchenko et al. \cite{grishchenko2018semantic} (${\cal D}_{27}$) define a complete small-step semantics of EVM bytecode and formalize most of the semantics in the proof assistant F* \cite{swamy2016dependent}. Grishchenko et al. \cite{grishchenko2018ethertrust, grishchenko2018foundations} (${\cal D}_{28}$)
leverage the complete small-step semantics of EVM bytecode to build {\it EtherTrust}, which is the first sound and automated static analyzer to achieve formal security related to the reachability properties of EVM bytecode. 
Other early-stage investigations include \cite{bhargavan2016formal,Why386:online,FormalVe73:online}.

\paragraph{Fuzzing}
Fuzz testing has been used to detect vulnerabilities in smart contracts. ContractFuzzer \cite{jiang2018contractfuzzer} (${\cal D}_{29}$) can detect five types of smart contract vulnerabilities, such as the {\em reentrancy} vulnerability ($\V_{1}$) and the {\em unchecked call return value} vulnerability ($\V_{15}$).
It generates inputs by crawling the ABI interfaces of smart contracts to extract their function selectors and data types of each argument, and instruments EVM to log contract execution behaviors for inspection. ReGuard \cite{liu2018reguard} (${\cal D}_{30}$) aims to detect the {\em reentrancy} vulnerability ($\V_{1}$) in smart contracts by transforming smart contracts to semantically equivalent C++ program and generating random transactions via a fuzzing engine to check the execution traces of the C++ program.

\paragraph{Model-based vulnerability detection}
Tann et al. \cite{tann2018towards} (${\cal D}_{31}$) use sequence learning to detect three types of vulnerabilities, namely the {\em frozen Ether} vulnerability ($\V_{3}$), the {\em leaking Ether to arbitrary address} vulnerability ($\V_{11}$), and
the {\em unprotected suicide} vulnerability ($\V_{10}$).
Tikhomirov et al. \cite{tikhomirov2018smartcheck} (${\cal D}_{32}$) propose SmartCheck to detect vulnerabilities in Solidity contracts by translating Solidity source code into an XML-based parse-tree and checks it against specific XPath patterns; they can detect 10 types of vulnerabilities, such as the {\em DoS with unexpected revert} vulnerability ($\V_{5}$), the {\em manipulated balance} vulnerability ($\V_{7}$), but incur a high false-positives rate.

\subsubsection{Contract and Ethereum enhancement}
Kosba et al. proposed the Hawk \cite{kosba2016hawk} (${\cal D}_{33}$) framework for incorporating cryptographic mechanisms to hide transaction data, effectively allowing contract developers to build privacy-preserving smart contracts without asking them to implement any cryptography. This framework can be used to defend against attacks that exploit the {\em secrecy failure} vulnerability ($\V_{12}$) and the {\em transaction-ordering dependence} vulnerability ($\V_{24}$). Zhang et al. \cite{zhang2016town} (${\cal D}_{34}$) designed an authenticated data feed system, dubbed Town Crier, for smart contracts that require access to external data. This system can be used to mitigate the {\em generating randomness} vulnerability ($\V_{26}$). Adler et al. \cite{adler2018astraea} (${\cal D}_{35}$) extended Town Crier by implementing a voting-based decentralized oracle to address the centralized point-of-failure that is inherent to Town Crier. Chen \cite{chen2017adaptive} (${\cal D}_{36}$) proposed an adaptive gas cost mechanism, which dynamically adjusts the cost of
EVM operations according to their execution times, so as to defend against DoS attacks that exploit the {\em under-priced opcodes} vulnerability ($\V_{23}$). In order to harden the Ethereum network against eclipse attacks, Marcus et al. \cite{marcus2018low} (${\cal D}_{37}$) proposed a set of countermeasures for eliminating some complicated artifacts of the Kademlia protocol used by Ethereum. Wang et al.  \cite{wang2018attack} (${\cal D}_{38}$) proposed some countermeasures for defending Ethereum clients against the attacks that may exploit the {\em RPC API exposure} vulnerability ($\V_{38}$).

\subsubsection{New blockchain protocols}
In order to tackle the {\em outsourceable puzzle} vulnerability ($\V_{29}$), Miller et al. \cite{miller2015nonoutsourceable} (${\cal D}_{39}$) formalized the notion of non-outsourceable puzzles and employed Merkle-tree-based proofs for puzzle design. The basic idea underlying  non-outsourceable puzzles is: If a pool operator outsources the mining task other miners, the miners can collect the full credit while the pool operator gets nothing, which effectively disincentivizes pool operators from outsourcing their mining tasks. Zeng et al. \cite{zeng2017nonoutsourceable} (${\cal D}_{40}$) extended the work of \cite{miller2015nonoutsourceable} by proposing a non-outsourceable puzzle that is compatible with the GHOST protocol used by Ethereum. Daian et al. \cite{daian2017short} (${\cal D}_{41}$) designed a two-stage non-outsourceable puzzle where the outer puzzle relies on the solution to the inner puzzle. 

In order to mitigate the {\em 51\% hashrate} vulnerability ($\V_{30}$), Eyal et al. \cite{HowtoDis46:online} (${\cal D}_{42}$) proposed the two-phase proof of work (2P-PoW) mechanism to disincentivize large mining pools, by incorporating two separate puzzles (instead of one) for miners to solve. Luu et al. \cite{luu2017smartpool} (${\cal D}_{43}$) implemented a new decentralized pooled mining protocol to defend against mining centralization, by replacing the traditional mining pool operator with an Ethereum smart contract.

\subsection{Reactive defenses}
Reactive defenses aim to react to potential exploitations of (unknown) vulnerabilities during the contract runtime to mitigate the damage.
A runtime verification method monitors the execution traces to detect and possibly react to suspicious activities that may violate certain properties. DappGuard \cite{cook2017dappguard} (${\cal D}_{44}$) actively monitors the incoming transactions to a smart contract and leverages the aforementioned tool Oyente \cite{luu2016making} to decide whether or not an incoming transaction can cause a security violation and if so, a counter transaction can be invoked to kill the contract in question. ContractLarva \cite{ellul2018runtime} (${\cal D}_{45}$) generates a new Solidity contract from the original contract and its specification, checks the original contract's runtime behaviors against this new contract's, and takes appropriate actions in the case of any discrepancy. 
Sereum \cite{rodler2018sereum} (${\cal D}_{46}$) uses
taint analysis to monitor runtime data flows during smart contract execution for detecting and preventing the {\em reentrancy} vulnerability ($\V_{1}$).

When detecting a violation, various mechanisms (${\cal D}_{47}$) have been proposed for mitigating the damage: (i) disabling the vulnerable smart contract or sensitive functionalities by using (for example) the {\it emergency stop} mechanism \cite{wohrer2018smart}; (ii) adopting a {\em stake-placing} mechanism to assure that any invocation, which potentially violates some properties, should pay a stake of compensation before running the contract and returns the stake back to the caller after the contract terminates normally; (iii) replacing vulnerable contracts with secure ones using the {\it virtual upgrade} mechanism \cite{software_engineering}, which can be realized by using a registry contract to hold the address of the latest version of a contract or by introducing a proxy contract to delegate calls to the latest version of a contract.

\subsection{Further analysis based on defense capabilities}

Now we present an analysis of defenses from the perspective of {\em defense capabilities}, meaning which defense can defend against the attacks that exploit certain vulnerabilities.

\begin{figure}[!htbp]
\includegraphics[width=.45\textwidth]{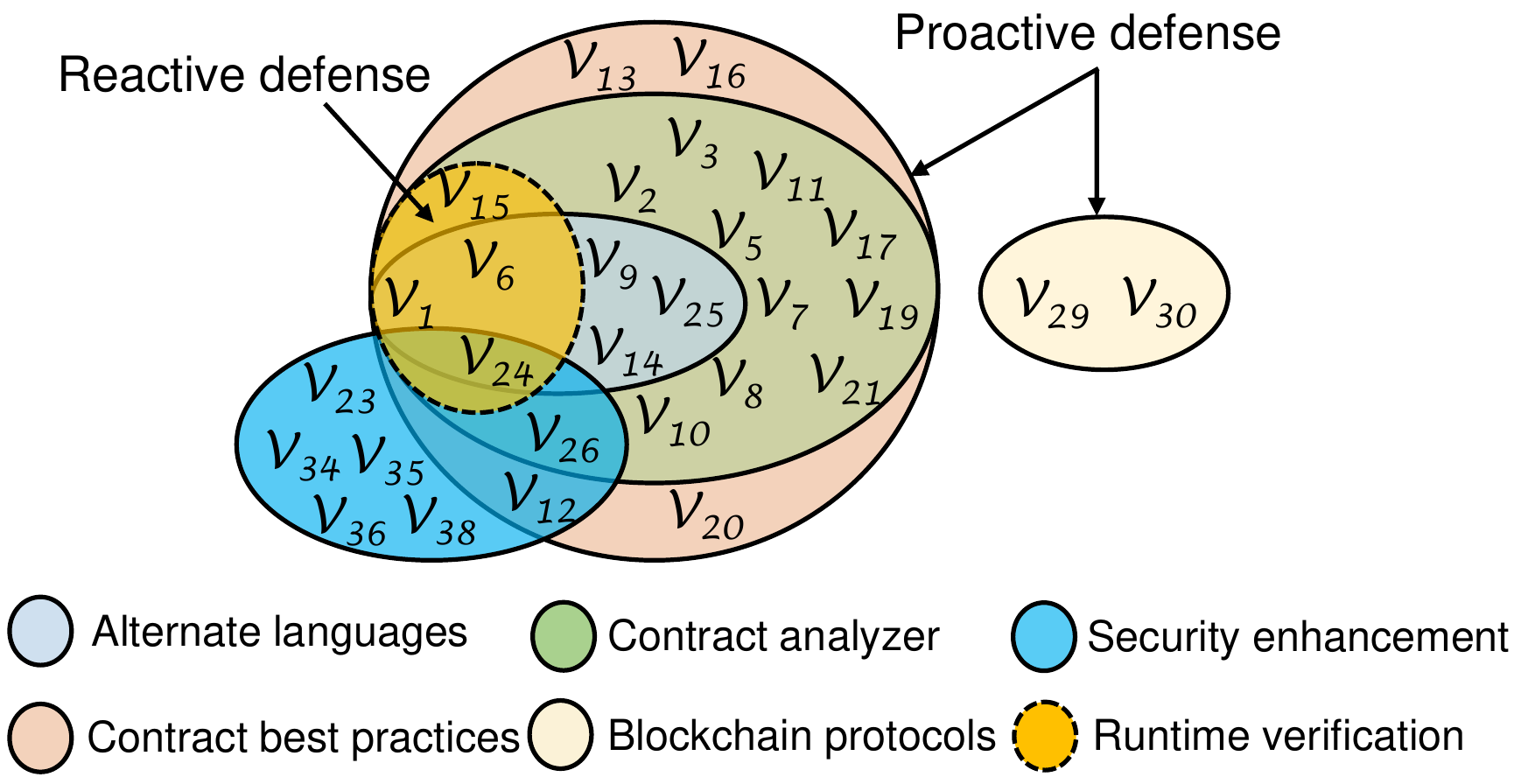}
\centering
\caption{Vern diagram representation of defenses against attacks that exploit some of 29 vulnerabilities; note that not all of the 44 vulnerabilities are relevant here because some vulnerabilities are already eliminated while other vulnerabilities are yet to be coped with.}
\label{fig:vuls-defense}
\end{figure}

Figure \ref{fig:vuls-defense} plots the Vern diagram of the 6 kinds of defenses discussed above, including 5 kinds of proactive defenses (i.e., {\em alternate language}, {\em contract analyzer}, {\em security enhancement}, {\em contract best practices}, {\em blockchain protocols}) and 1 kind of reactive defenses (i.e., {\em runtime verification}).
For proactive defense, we observe the following: (i) using {\em contract best practices} in the course of developing contracts can prevent or mitigate attacks that attempt to exploit 22 types of application-layer vulnerabilities; 
(ii) using {\em smart contract analyzer} can detect or mitigate attacks that attempts to exploit 18 types of application-layer vulnerabilities; 
(iii) using {\em security enhancement} can prevent or mitigate attacks that attempts to exploit 8 types of vulnerabilities, including 4 application-layer vulnerabilities and 4 network-layer ones; 
(iv) using better-designed contract programming language (i.e., {\em alternate languages} for short) can prevent or mitigate attacks that attempt to exploit 6 types of application-layer vulnerabilities; 
(v) using better-designed {\em blockchain protocols} can mitigate attacks that attempt to exploit 2 types of consensus-layer vulnerabilities. For reactive defense, we observe that using {\em runtime verification} can mitigate attacks that attempt to exploit 4 types of application-layer vulnerabilities.
In total, proactive defenses can defend against attacks that attempt to exploit 29 types of vulnerabilities, whereas reactive defenses can defend attacks that attempt to exploit 4 types of vulnerabilities, which are also covered by proactive defenses. Moreover, we observe that the vulnerabilities that can be defended by {\em alternate languages} and {\em runtime verification} are subsets of the vulnerabilities that can be defended by {\em contract analyzers}, which are in turn a subset of the vulnerabilities that can be defended by {\em contract development best practices}. 
We draw the following insights.

\begin{insight}
\label{insight:best-practices}
Industry has come up with a significant set of best practices for guiding the development of smart contracts.
\end{insight}

Insight \ref{insight:best-practices} highlights the importance of adopting best practices in the process of developing software. Nevertheless, these best practices alone are not adequate in assuring security.

\begin{insight}
\label{insight:proactive-vs-reactive}
Proactive defenses can defend against attacks that attempt to exploit many vulnerabilities. In contrast, reactive defenses can defend against attacks that attempt to exploit a few vulnerabilities.
\end{insight}

Insight \ref{insight:proactive-vs-reactive} reflects the state-of-the-art. Nevertheless, reactive defenses are still important because they may be able to defend attacks that attempt to exploit vulnerabilities that survived proactive defenses.

\subsection{Further analysis based on defense investment}
Now we present an analysis of defenses from the perspective of {\em defense investment}, meaning how much effort has been invested in designing defense against attacks that exploit a certain vulnerability. 
However, we note that some defenses are {\em not} geared towards any specific vulnerabilities or attacks; for example, software engineering mechanisms (i.e., ${\cal D}_{8}$) are neither geared towards any specific vulnerability nor geared towards any specific attack.

\begin{figure}[!htbp]
\includegraphics[width=.45\textwidth]{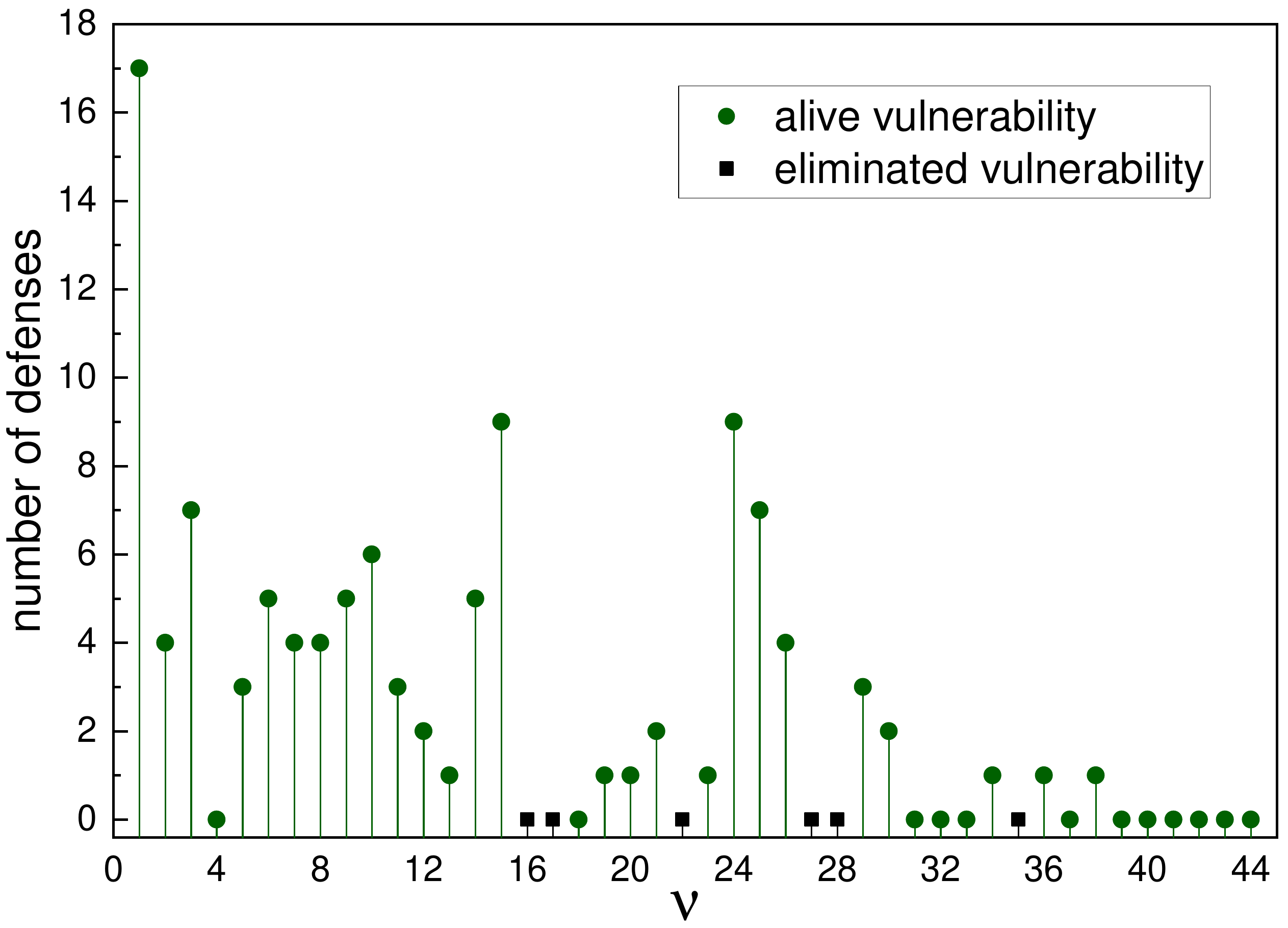}
\centering
\caption{The number defenses with respect to each vulnerability.}
\label{fig:vuls-defense2}
\end{figure}

\renewcommand{\arraystretch}{0.6}
\begin{table*}[!htbp]
  \centering
  \scriptsize
  \setlength\tabcolsep{0.6pt}
  %\arraystretch{0.5pt}
  \caption{Elaboration of the defenses with respect to vulnerabilities summarized in Figure \ref{fig:vuls-defense2}.}
    \begin{tabular}{|c|c|c|c|c|c|c|c|c|c|c|c|c|c|c|c|c|c|c|c|c|c|c|c|c|c|c|c|c|c|c|c|c|c|c|c|c|c|c|c|c|c|c|c|c|c|}
    \hline
          & ${\cal V}_{1}$    & ${\cal V}_{2}$    & ${\cal V}_{3}$    & ${\cal V}_{4}$    & ${\cal V}_{5}$    & ${\cal V}_{6}$    & ${\cal V}_{7}$    & ${\cal V}_{8}$    & ${\cal V}_{9}$    & ${\cal V}_{10}$   & ${\cal V}_{11}$   & ${\cal V}_{12}$   & ${\cal V}_{13}$   & ${\cal V}_{14}$   & ${\cal V}_{15}$   & ${\cal V}_{16}$   & ${\cal V}_{17}$   & ${\cal V}_{18}$   & ${\cal V}_{19}$   & ${\cal V}_{20}$   & ${\cal V}_{21}$   & ${\cal V}_{22}$   & ${\cal V}_{23}$   & ${\cal V}_{24}$   & ${\cal V}_{25}$   & ${\cal V}_{26}$   & ${\cal V}_{27}$   & ${\cal V}_{28}$   & ${\cal V}_{29}$   & ${\cal V}_{30}$   & ${\cal V}_{31}$   & ${\cal V}_{32}$   & ${\cal V}_{33}$   & ${\cal V}_{34}$   & ${\cal V}_{35}$   & ${\cal V}_{36}$   & ${\cal V}_{37}$   & ${\cal V}_{38}$   & ${\cal V}_{39}$   & ${\cal V}_{40}$   & ${\cal V}_{41}$   & ${\cal V}_{42}$   & ${\cal V}_{43}$   & ${\cal V}_{44}$   & total \bigstrut\\
    \hline
    ${\cal D}_{1}$    &       &       &       &       &       &  \footnotesize{\checkmark}     &       &       & \footnotesize{\checkmark}     &       &       &       &       & \footnotesize{\checkmark}     &       &      &       &       &       &       &       &       &       &       &       &       &       &       &       &       &       &       &       &       &       &       &       &       &       &       &       &       &       &       & 3 \bigstrut\\
    \hline
    ${\cal D}_{2}$    & \footnotesize{\checkmark}     &       &       &       &       &       &       &       &       &       &       &       &       & \footnotesize{\checkmark}     &       &       &       &       &       &       &       &       &       & \footnotesize{\checkmark}     &       &       &       &       &       &       &       &       &       &       &       &       &       &       &       &       &       &       &       &       & 3 \bigstrut\\
    \hline
    ${\cal D}_{3}$    & \footnotesize{\checkmark}     &       &       &       &       &       &       &       &       &       &       &       &       &       &       &       &       &       &       &       &       &       &       &       &       &       &       &       &       &       &       &       &       &       &       &       &       &       &       &       &       &       &       &       & 1 \bigstrut\\
    \hline
    ${\cal D}_{4}$    & \footnotesize{\checkmark}     &       &       &       &       & \footnotesize{\checkmark}     &       &       & \footnotesize{\checkmark}     &       &       &       &       &       &       &       &       &       &       &       &       &       &       &       & \footnotesize{\checkmark}     &       &       &       &       &       &       &       &       &       &       &       &       &       &       &       &       &       &       &       & 4 \bigstrut\\
    \hline
    ${\cal D}_{7}$    & \footnotesize{\checkmark}     & \footnotesize{\checkmark}     & \footnotesize{\checkmark}     &       & \footnotesize{\checkmark}     & \footnotesize{\checkmark}     & \footnotesize{\checkmark}     & \footnotesize{\checkmark}     & \footnotesize{\checkmark}     & \footnotesize{\checkmark}     & \footnotesize{\checkmark}     & \footnotesize{\checkmark}     & \footnotesize{\checkmark}     & \footnotesize{\checkmark}     & \footnotesize{\checkmark}     & \footnotesize{\checkmark}     & \footnotesize{\checkmark}     &       & \footnotesize{\checkmark}     & \footnotesize{\checkmark}     & \footnotesize{\checkmark}     &       &       & \footnotesize{\checkmark}     & \footnotesize{\checkmark}     & \footnotesize{\checkmark}     &       &       &       &       &       &       &       &       &       &       &       &       &       &       &       &       &       &       & 22 \bigstrut\\
    \hline
    ${\cal D}_{9}$    & \footnotesize{\checkmark}     &       &       &       &       &       &       &       &       &       &       &       &       &       & \footnotesize{\checkmark}     &       &       &       &       &       &       &       &       & \footnotesize{\checkmark}     & \footnotesize{\checkmark}     &       &       &       &       &       &       &       &       &       &       &       &       &       &       &       &       &       &       &       & 4 \bigstrut\\
    \hline
    ${\cal D}_{10}$   &       &       & \footnotesize{\checkmark}     &       &       &       &       &       &       & \footnotesize{\checkmark}     & \footnotesize{\checkmark}     &       &       &       &       &       &       &       &       &       &       &       &       &       &       &       &       &       &       &       &       &       &       &       &       &       &       &       &       &       &       &       &       &       & 3 \bigstrut\\
    \hline
    ${\cal D}_{11}$   & \footnotesize{\checkmark}      &       &      &       &       &       &   \footnotesize{\checkmark}    &   \footnotesize{\checkmark}    &       & \footnotesize{\checkmark}     &     &       &       &       &  \footnotesize{\checkmark}     &       &       &       &       &       &       &       &       &  \footnotesize{\checkmark}     &    \footnotesize{\checkmark}   &   \footnotesize{\checkmark}    &       &       &       &       &       &       &       &       &       &       &       &       &       &       &       &       &       &       & 8 \bigstrut\\
    \hline    
    ${\cal D}_{12}$   &       &       &       &       &       &       &       &       & \footnotesize{\checkmark}     &       &       &       &       &       &       &       & \footnotesize{\checkmark}     &       &       &       &       &       &       &       &       &       &       &       &       &       &       &       &       &       &       &       &       &       &       &       &       &       &       &       & 2 \bigstrut\\
    \hline
    ${\cal D}_{13}$   & \footnotesize{\checkmark}     &       & \footnotesize{\checkmark}     &       &       &       &       &       &       &  \footnotesize{\checkmark}     &       &       &       &       &       &       &       &       &       &       & \footnotesize{\checkmark}     &       &       &       &       &       &       &       &       &       &       &       &       &       &       &       &       &       &       &       &       &       &       &       & 4 \bigstrut\\
    \hline
    ${\cal D}_{14}$   &\footnotesize{\checkmark}     &       &       &       &       &       &       &       &       &       &       &       &       &       &       &       &       &       &       &       &       &       &       &       &       &       &       &       &       &       &       &       &       &       &       &       &       &       &       &       &       &       &       &       & 1 \bigstrut\\
    \hline
    ${\cal D}_{15}$   & \footnotesize{\checkmark}     & \footnotesize{\checkmark}     & \footnotesize{\checkmark}     &       &       &       &       &       &       &       &       &       &       &       & \footnotesize{\checkmark}     &       &       &       &       &       &       &       &       & \footnotesize{\checkmark}     &       &       &       &       &       &       &       &       &       &       &       &       &       &       &       &       &       &       &       &       & 5 \bigstrut\\
    \hline
    ${\cal D}_{16}$   & \footnotesize{\checkmark}     &       &       &       &    \footnotesize{\checkmark}   & \footnotesize{\checkmark}     &       &       &       &       &       &       &       &       & \footnotesize{\checkmark}     &       &       &       &       &       &       &       &       & \footnotesize{\checkmark}     & \footnotesize{\checkmark}     &       &       &       &       &       &       &       &       &       &       &       &       &       &       &       &       &       &       &       & 6 \bigstrut\\
    \hline
    ${\cal D}_{17}$   & \footnotesize{\checkmark}     &       &       &       &       &       &       &       &       &       &       &       &       &       &       &       &       &       &       &       &       &       &       &\footnotesize{\checkmark}     &       &       &       &       &       &       &       &       &       &       &       &       &       &       &       &       &       &       &       &       & 2 \bigstrut\\
    \hline
    ${\cal D}_{18}$   &       &       &       &       &       &       &       &       &       &       &       &       &       & \footnotesize{\checkmark}     &       &       &       &       &       &       &       &       &       &       &       &       &       &       &       &       &       &       &       &       &       &       &       &       &       &       &       &       &       &       & 1 \bigstrut\\
    \hline
    ${\cal D}_{19}$   & \footnotesize{\checkmark}     &       &       &       &       &       &  \footnotesize{\checkmark}      & \footnotesize{\checkmark}     &       & \footnotesize{\checkmark}     &       &       &       &       & \footnotesize{\checkmark}     &       &       &      &       &       &       &       &       &       &       &       &       &       &       &       &       &       &       &       &       &       &       &       &       &       &       &       &       &       & 5 \bigstrut\\
    \hline
    ${\cal D}_{29}$   & \footnotesize{\checkmark}     & \footnotesize{\checkmark}     & \footnotesize{\checkmark}     &       &       &       &       &       &       &       &       &       &       &       & \footnotesize{\checkmark}     &       &       &       &       &       &       &       &       &       & \footnotesize{\checkmark}     &       &       &       &       &       &       &       &       &       &       &       &       &       &       &       &       &       &       &       & 5 \bigstrut\\
    \hline
    ${\cal D}_{30}$   & \footnotesize{\checkmark}     &       &       &       &       &       &       &       &       &       &       &       &       &       &       &       &       &       &       &       &       &       &       &       &       &       &       &       &       &       &       &       &       &       &       &       &       &       &       &       &       &       &       &       & 1 \bigstrut\\
    \hline
    ${\cal D}_{31}$   &       &       & \footnotesize{\checkmark}     &       &       &       &       &       &       & \footnotesize{\checkmark}     & \footnotesize{\checkmark}     &       &       &       &       &       &       &       &       &       &       &       &       &       &       &       &       &       &       &       &       &       &       &       &       &       &       &       &       &       &       &       &       &       & 3 \bigstrut\\
    \hline
    ${\cal D}_{32}$   & \footnotesize{\checkmark}     &   \footnotesize{\checkmark}    & \footnotesize{\checkmark}     &       & \footnotesize{\checkmark}     &       & \footnotesize{\checkmark}     & \footnotesize{\checkmark}     & \footnotesize{\checkmark}     &       &       &       &       & \footnotesize{\checkmark}     & \footnotesize{\checkmark}     &       &       &       &     &       &       &       &       &       & \footnotesize{\checkmark}     &       &       &       &       &       &       &       &       &       &       &       &       &       &       &       &       &       &       &       & 10 \bigstrut\\
    \hline
    ${\cal D}_{33}$   &       &       &       &       &       &       &       &       &       &       &       & \footnotesize{\checkmark}     &       &       &       &       &       &       &       &       &       &       &       & \footnotesize{\checkmark}     &       &       &       &       &       &       &       &       &       &       &       &       &       &       &       &       &       &       &       &       & 2 \bigstrut\\
    \hline
    ${\cal D}_{34}$   &       &       &       &       &       &       &       &       &       &       &       &       &       &       &       &       &       &       &       &       &       &       &       &       &       & \footnotesize{\checkmark}     &       &       &       &       &       &       &       &       &       &       &       &       &       &       &       &       &       &       & 1 \bigstrut\\
    \hline
    ${\cal D}_{35}$   &       &       &       &       &       &       &       &       &       &       &       &       &       &       &       &       &       &       &       &       &       &       &       &       &       & \footnotesize{\checkmark}     &       &       &       &       &       &       &       &       &       &       &       &       &       &       &       &       &       &       & 1 \bigstrut\\
    \hline
    ${\cal D}_{36}$   &       &       &       &       &       &       &       &       &       &       &       &       &       &       &       &       &       &       &       &       &       &       & \footnotesize{\checkmark}     &       &       &       &       &       &       &       &       &       &       &       &       &       &       &       &       &       &       &       &       &       & 1 \bigstrut\\
    \hline
    ${\cal D}_{37}$   &       &       &       &       &       &       &       &       &       &       &       &       &       &       &       &       &       &       &       &       &       &       &       &       &       &       &       &       &       &       &       &       &       & \footnotesize{\checkmark}     & \footnotesize{\checkmark}     & \footnotesize{\checkmark}     &       &       &       &       &       &       &       &       & 3 \bigstrut\\
    \hline
    ${\cal D}_{38}$   &       &       &       &       &       &       &       &       &       &       &       &       &       &       &       &       &       &       &       &       &       &       &       &       &       &       &       &       &       &       &       &       &       &       &       &       &       & \footnotesize{\checkmark}     &       &       &       &       &       &       & 1 \bigstrut\\
    \hline
    ${\cal D}_{39}$   &       &       &       &       &       &       &       &       &       &       &       &       &       &       &       &       &       &       &       &       &       &       &       &       &       &       &       &       & \footnotesize{\checkmark}     &       &       &       &       &       &       &       &       &       &       &       &       &       &       &       & 1 \bigstrut\\
    \hline
    ${\cal D}_{40}$   &       &       &       &       &       &       &       &       &       &       &       &       &       &       &       &       &       &       &       &       &       &       &       &       &       &       &       &       & \footnotesize{\checkmark}     &       &       &       &       &       &       &       &       &       &       &       &       &       &       &       & 1 \bigstrut\\
    \hline
    ${\cal D}_{41}$   &       &       &       &       &       &       &       &       &       &       &       &       &       &       &       &       &       &       &       &       &       &       &       &       &       &       &       &       & \footnotesize{\checkmark}     &       &       &       &       &       &       &       &       &       &       &       &       &       &       &       & 1 \bigstrut\\
    \hline
    ${\cal D}_{42}$   &       &       &       &       &       &       &       &       &       &       &       &       &       &       &       &       &       &       &       &       &       &       &       &       &       &       &       &       &       & \footnotesize{\checkmark}     &       &       &       &       &       &       &       &       &       &       &       &       &       &       & 1 \bigstrut\\
    \hline
    ${\cal D}_{43}$   &       &       &       &       &       &       &       &       &       &       &       &       &       &       &       &       &       &       &       &       &       &       &       &       &       &       &       &       &       & \footnotesize{\checkmark}     &       &       &       &       &       &       &       &       &       &       &       &       &       &       & 1 \bigstrut\\
    \hline
    ${\cal D}_{44}$   & \footnotesize{\checkmark}     &       &       &       &       & \footnotesize{\checkmark}     &       &       &       &       &       &       &       &       & \footnotesize{\checkmark}     &       &       &       &       &       &       &       &       & \footnotesize{\checkmark}     &       &       &       &       &       &       &       &       &       &       &       &       &       &       &       &       &       &       &       &       & 4 \bigstrut\\
    \hline
    ${\cal D}_{46}$   & \footnotesize{\checkmark}     &       &       &       &       &       &       &       &       &       &       &       &       &       &       &       &       &       &       &       &       &       &       &       &       &       &       &       &       &       &       &       &       &       &       &       &       &       &       &       &       &       &       &       & 1 \bigstrut\\
    \hline
    total & 17    & 4     & 7     & 0     & 3     & 5     & 4     & 4     & 5     & 6     & 3     & 2     & 1     & 5     & 9     & --     & --     & 0     & 1     & 1     & 2     & --     & 1     & 9     & 7     & 4     & --     & --     & 3     & 2     & 0     & 0     & 0     & 1     & --     & 1     & 0     & 1     & 0     & 0     & 0     & 0     & 0     & 0     &  \bigstrut\\
    \hline
    \end{tabular}%
  \label{tab:elaboration-of-defenses-vs-vulnerabilities}%
\end{table*}%

Figure \ref{fig:vuls-defense2} plots the summary of the number of defense mechanisms with respect to individual vulnerabilities. Table  \ref{tab:elaboration-of-defenses-vs-vulnerabilities} elaborates which vulnerabilities may be protected by which defenses. For example, Figure \ref{fig:vuls-defense2} shows that $\V_1$ can be defended by 17 defenses; Table  \ref{tab:elaboration-of-defenses-vs-vulnerabilities} shows which these 17 defenses are.
On one hand, we observe that the most extensively investigated vulnerability is the {\em reentrancy} vulnerability (${\cal V}_1$), which can be mitigated by 17 kinds of proactive and reactive defense mechanisms. Other vulnerabilities that have been substantially investigated include: the {\em unchecked call return value} vulnerability (${\cal V}_{15}$) and the {\em transaction-ordering dependence} vulnerability (${\cal V}_{24}$), each of which can be defended by 9 kinds of proactive and reactive defense mechanisms; the {\em frozen Ether} vulnerability (${\cal V}_3$) and the {\em timestamp dependence} vulnerability (${\cal V}_{25}$) can be defended by 7 kinds of proactive defense mechanisms. It appears that the vulnerabilities that have been thoroughly investigated are (i) the ones that have caused large financial losses, (ii) the ones that are inherent to the design of the Solidity language, and (iii) the ones that are inherent to the profit-making factor for assembling blocks. This is so because the {\em reentrancy} vulnerability (${\cal V}_1$) has caused the loss of US\$60M in the DAO attack (i.e., attack ${\cal A}_1$ in Figure \ref{fig:vuls-attack-consequence}), the {\em frozen Ether} vulnerability (${\cal V}_3$) has caused the loss of US\$280M in the Parity wallet attack (i.e., attack ${\cal A}_3$), the {\em unchecked call return value} vulnerability (${\cal V}_{15}$) is inherent to the exception handling mechanism in the Solidity language, the {\em transaction-ordering dependence} vulnerability (${\cal V}_{24}$) is inherent to the unpredictable nature of the Ethereum blockchain, and the {\em timestamp dependence} vulnerability (${\cal V}_{25}$) is inherent to the manipulable block information of the Ethereum blockchain. On the other hand, we observe 19 vulnerabilities that have zero or one defense mechanism. Most of these vulnerabilities are either caused by the Ethereum design and implementation, or reside in the Ethereum environment.
This leads to:

\begin{insight}
\label{insight:defense-effort-discrepancy}
There is a large discrepancy between the effort that has been invested to defend against attacks that exploit different vulnerabilities.
\end{insight}

Insight \ref{insight:defense-effort-discrepancy}
is interesting because it seems that the defense effort has been driven by the consequential financial loss incurred by the exploitation of certain vulnerabilities. This prioritization strategy is {\em not} adequate because it suggests in a sense that the defender is always chasing behind the attacker, who detects and exploits an vulnerability that then becomes known to the defender. 

\begin{insight}
\label{insight:defense-weakness}
Existing studies focus on defending against attacks that attempt to exploit
vulnerabilities in the DApp back-end (i.e., smart contracts), but largely ignore the protection of the DApp front-end (i.e., browser) and the interactions between the front-end and the back-end.
\end{insight}

Insight \ref{insight:defense-weakness} says that 
in order to adequately defend DApps, a good solution should consider the front-end interface, the back-end contracts, and their interactions.
It is worth mentioning that attacks  ${\cal A}_{23}$, ${\cal A}_{24}$, ${\cal A}_{25}$, and ${\cal A}_{26}$ have exploited either the front-end interface or the front-end and back-end interactions to cause the loss of over US\$24M in total.

\section{Discussion: Going Beyond Ethereum}
\label{sec:further-discussion}

The preceding discussion is geared towards the Ethereum system and offers a number of open problems for future research. While the findings and (some of) the open problems might be applicable to blockchain-based systems in general, in what follows we discuss future research directions that are equally applicable to Ethereum and blockchain-based systems.

\subsection{Rigorous definition of properties}

We observe that in order to adequately defend Ethereum and blockchain-based systems, there is an urgent need to understand the desirable security properties, which are however extremely difficult to formalize for a  complex system like Ethereum. Some informal properties have been discussed in \cite{DBLP:journals/corr/abs-1806-04358}, which represents a very preliminary first step towards the ultimate goal. This leads to:

\begin{insight}
There is a lack of deep understanding on the rigorously-specified desirable security properties that should be possessed by Ethereum and blockchain-based systems.
\end{insight}

\subsection{Rigorous analysis methodologies}

Having defined the rigorous properties that should be satisfied by Ethereum and blockchain-based systems, we need principled and rigorous methodologies to analyze that the desirable properties are indeed satisfied. For properties that are geared towards building-blocks, cryptography and formal methods have been two successful approaches, while noting that they have their own limitations. Since blockchain-based systems, such as Ethereum, as indeed complex systems, it is probably infeasible to prevent all attacks, meaning that attacks are inevitable and their risks must be adequately understood manageable. This calls for rigorous analysis methodologies from a holistic perspective. Towards this ultimate goal, the recently proposed approach of {\em cybersecurity dynamics} 
\cite{XuCybersecurityDynamicsHotSoS2014,XuEmergentBehaviorHotSoS2014,XuBookChapterCD2019,
XuTAAS2012,
XuHotSOS14-MTD,XuHotSoS2015,XuInternetMath2015Dependence,XuIEEETNSE2018,XuIEEEACMToN22019} has a great potential, although we are not aware of any results that have been published in the literature. This leads to:
\begin{insight}
There is a lack of deep understanding on the rigorous analysis methodologies that are necessary and sufficient for analyzing the desired properties of blockchain-based systems.
\end{insight}

\subsection{Metrics}

It is a notoriously difficult problem to rigorously define metrics to systematically measure security properties of interest. However, given the high likelihood that blockchain-based systems will become digital 
infrastructures for the future society, if not already, there is an urgent need to define metrics to measure their security and risk. We are not aware of any systematic study on this matter, despite substantial efforts 
\cite{DBLP:journals/tdsc/NicolST04,Pendleton16,8017389,Noel2017,
XuHotSoS2018Firewall,
XuAgility2019,XuSTRAM2018ACMCSUR}. This leads to:
\begin{insight}
There is a lack of deep understanding on the metrics that are necessary and sufficient for quantifying the security and risk of blockchain-based systems.
\end{insight}

\section{Conclusion}
\label{sec:conclusion}
We have presented a systematic survey on the security of the Ethereum system, including its application, data, consensus, and network layers. The survey considered three perspectives, namely vulnerabilities, attacks and defenses, while correlating them. We discussed not only the locations of the vulnerabilities, but also their root causes. We systematized the attacks against, and the defenses for, the Ethereum system. We further systematized the best practices proposed by industry into a small number of guiding principles, which might be easier to adopt by practitioners. We provide insights into the state-of-the-art and into future research directions.

\smallskip

\noindent{\bf Acknowledgement}. Approved for  Public  Release;  Distribution  Unlimited:  88ABW-2019-3890 Dated 12 August 2019.

\Urlmuskip=0mu plus 1mu\relax
\bibliographystyle{ieeetr}
%\bibliography{ref}

\end{document}